\documentclass[prb,superscriptaddress,amsfonts,amssymb,amsmath,floats,twocolumn,aps,footinbib,shownopacs]{revtex4-2}

\usepackage{graphicx}
\usepackage{dcolumn}
\usepackage{bm}
\usepackage[pdftex,colorlinks=true, linkcolor=blue,citecolor=blue,filecolor=blue]{hyperref}

\usepackage{xcolor}

\begin{document}
\preprint{APS/123-QED}
	
\title{Dynamical mean-field study of a photon-mediated ferroelectric phase transition}
	
\author{Katharina Lenk}
\affiliation{Department of Physics, University of Erlangen-N\"urnberg, 91058 Erlangen, Germany}
\author{Jiajun Li}
\affiliation{Department of Physics, University of Fribourg, 1700 Fribourg Switzerland}
\affiliation{Paul Scherrer Institute, Condensed Matter Theory, PSI Villigen, Switzerland}
\author{Philipp Werner}
\affiliation{Department of Physics, University of Fribourg, 1700 Fribourg Switzerland}
\author{Martin Eckstein}
\affiliation{Department of Physics, University of Erlangen-N\"urnberg, 91058 Erlangen, Germany}

\date{\today}
	
\begin{abstract}
The interplay of light and matter gives rise to intriguing cooperative effects in quantum many-body systems. This is even true in thermal equilibrium, where 
the electromagnetic field can hybridize  with collective modes of matter, and 
virtual photons can induce interactions in the solid. Here, we show how these light-mediated interactions can be treated using the dynamical mean-field theory formalism. We consider a 
minimal
model of a two-dimensional material that couples to a surface plasmon polariton mode of a metal-dielectric interface. Within the mean-field approximation, the system exhibits a ferroelectric phase transition that is unaffected by the light-matter coupling. Bosonic dynamical mean-field theory provides a more accurate description and reveals that the photon-mediated interactions enhance the ferroelectric order and stabilize the ferroelectric phase.
\end{abstract}

\maketitle
	
\section{Introduction} 
	
The interplay of light and matter can lead to dramatic changes in the behavior of a system, which provides intriguing pathways to manipulate complex materials out of equilibrium \cite{Basov2017,Giannetti2016,delaTorre2021}. 
Even without external driving
the electromagnetic field can influence the properties of matter, 
by hybridizing 
with electromagnetically active modes in the solid, and through photon-mediated interactions. In free space, such effects are usually negligible.  In cavity or waveguide quantum electrodynamics, however, ultra-strong  coupling of photons  to individual emitters or molecules can be achieved by  structuring the electromagnetic field \cite{FriskKockum2019}, which may even be used to manipulate chemical reactions \cite{Flick2017, Fregoni2022, Schaefer2019}.   An intriguing question is therefore whether similar techniques can be used to manipulate  the collective behavior and  phase transitions in extended condensed matter systems \cite{Schlawin2022, Scalari2012, Zhang2016,Thomas2019,Thomas2021,Appugliese2022}. Theoretical proposals along these lines include photon-mediated and enhanced superconductivity \cite{Sentef2018,Schlawin2019, Li2020b}, photon-induced topological phases \cite{Wang2019,Dmytruk2022}, or the control of Mott polaritons and magnetic exchange interactions \cite{Kiffner2019,Sentef2020,MuellerH2022, Curtis2022}, spin liquids \cite{Chiocchetta2021}, and various forms of ferroelectricity \cite{Mazza2019, Lenk2020, Ashida2020, Latini2021, Schuler2020, Bernardis2018,Keeling2007}.

In an extended solid, one would expect that the effect of light on the thermodynamic properties of matter arises not from the coupling to a single cavity mode \cite{Lenk2022,Pilar2020}, but from a mode continuum. In a coplanar cavity setting, e.g., interactions within 
matter are mediated by 
photons
with a continuous  in-plane momentum. Because of the large light velocity, such a cavity would however affect only modes in an extremely small phase space volume compared to the Brillouin zone of the solid; thus, 
in condensed matter settings, it is often sufficient to 
consider the direct screened Coulomb interaction and neglect the effect of the transverse electromagnetic field (photon field).  A promising pathway to overcome this phase space constraint includes geometries in which propagating modes are strongly localized at interfaces, such as surface plasmons \cite{Ashida2020}, or possibly the polaritons of hyperbolic materials \cite{Caldwell2014}. The nanoscale confinement allows such modes to efficiently couple to a macroscopic number of degrees of freedom of a thin layer or a two dimensional material close to the interface. 

In this work we investigate within a microscopic model how the coupling to a surface plasmon mode can affect a ferroelectric transition in a  layered material. Specifically, we consider a minimal model of a two-dimensional material that exhibits 
a paraelectric to ferroelectric phase transition similar to the case of SrTiO$_3$ \cite{Mueller1979}, with a quantum paraelectric phase at zero temperature in a certain parameter regime.  
The coupling to the surface plasmon mode induces a long-range interaction in the material, which can be controlled by the distance of the material from the interface. With the  coupling to a mode continuum,  the system  constitutes a complex quantum many-body problem which cannot be solved exactly.  Apart from that, a conventional static mean-field approach, which often provides a reasonable starting point to get a qualitative understanding of a phase transition, does not capture the effect of the transverse field on the transition in the present case: The mean-field approximation replaces the induced interaction by a static and uniform self-consistent field; this mean-field, however, vanishes, because the  interactions  between dipoles that are induced by the transverse field are zero in the static limit due to the positive definite light-matter coupling. 
This fact also underlies the no-go theorems which exclude the condensation of hybrid light-matter modes due to the coupling to a single cavity mode \cite{Lenk2020,Andolina2019,Andolina2020,Lenk2022}.

On the phenomenological level, an effect of the electromagnetic field on the static thermodynamic properties beyond the mean-field limit can arise from interactions
in the solid: 
The transverse photon field can influence  electromagnetically active modes in the solid at frequency $\omega>0$, which in turn renormalize the static ($\omega=0$) response through anharmonic interactions between the modes \cite{Ashida2020}. Moreover, at temperature $T>0$ the total free energy content of all modes changes when they hybridize with the electromagnetic field. To establish a microscopic description of the system beyond static mean-field theory, we set up a solution within dynamical mean-field theory (DMFT).  
DMFT,
which becomes exact in the limit of large coordination number \cite{Metzner1989}, is commonly used in condensed matter 
physics, and 
one of the most versatile techniques for studying systems with strongly correlated electrons \cite{Georges1996}. To study a system with strong light-matter coupling, we apply the idea of DMFT to the fields that mediate the interaction, along the lines of an extended DMFT formalism \cite{Sun2002}. In the present case, where we consider an ionic solid without itinerant electrons, the formalism then becomes similar to bosonic DMFT \cite{Byczuk2008, Hu2009, Anders2010,Anders2011, Akerlund2013, Akerlund2014}. DMFT is an embedding approach, which maps a lattice model with local interaction to a self-consistent impurity model. For the light-matter coupled system, this impurity model is a generalized spin-boson model with a self-consistently determined continuum of bosonic modes. Depending on the parameters of the lattice model, this model can be in the ultra-strong coupling regime; in fact, although the impurity model is an auxiliary system, it provides a useful way to quantify whether a single site in the solid is effectively driven into the single-particle strong coupling regime. We solve this  impurity model numerically using a recently developed systematic diagrammatic many-body approach \cite{Aaram2021, Aaram2022}, which allows to deal with emitters coupled to a continuum, as needed in nonperturbative waveguide quantum electrodynamics \cite{Ashida2021b}.

The remainder of this paper is structured as follows: In Sec.~\ref{sec:model}, we introduce the total Hamiltonian of the system and derive the effective action for the material. In Sec.~\ref{sec:BDMFT} we explain the bosonic DMFT formalism, and briefly discuss the numerical implementation. The results of the calculations are presented in Sec.~\ref{sec:results}, and in Sec. \ref{sec:conclusion} we conclude 
with
a brief discussion and summary.

\section{Model}
\label{sec:model}
	
\begin{figure}
\centering
\includegraphics[width=\linewidth]{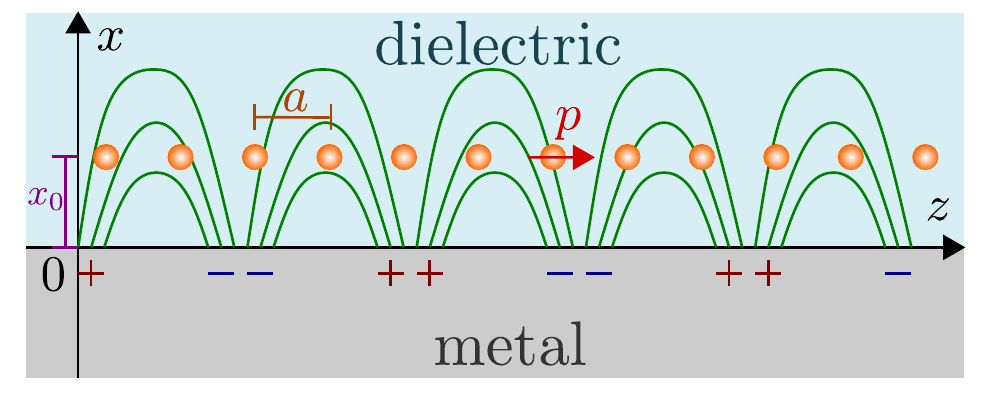}
\caption{Two-dimensional slice through the setting in the $x$-$z$-plane. The system consists of a metal dielectric interface and a collection of dipoles (orange dots) with dipole moment $p$ arranged on a square lattice of lattice constant $a$. The green lines represent the electric field lines of a SPP mode propagating in the $z$-direction.}
\label{fig:model}
\end{figure}

We consider a simple hetero-structure consisting of a two-dimensional material that is placed parallel to a metallic surface in the $y$-$z$-plane (see Fig.~\ref{fig:model} for a schematic sketch of the setup). The material consists of a collection of dipoles arranged on a square lattice. It is embedded in a dielectric medium of relative permittivity $\epsilon$, and located at a fixed distance $x_0$ from the metal-dielectric interface. The metal surface supports a propagating electromagnetic mode that gives rise to an evanescent field in both the metallic and the dielectric region. If the material is close enough to the surface, it couples to this so called surface plasmon polariton (SPP) mode. Due to the strong confinement of the electromagnetic field to the metal-dielectric interface, the coupling between light and matter is enhanced as compared to the vacuum case. 
Instead, the modes of a coplanar cavity with photon dispersion $\omega_{q}=\sqrt{(cq)^2+\omega_0^2}$ (momentum $q$ along the plane, fundamental cavity frequency $\omega_0$), which are extended over the full transverse cavity volume, would have a negligible effect on the 
material (see Appendix).

\subsection{Hamiltonian description}

We start from a Hamiltonian description of the system. 
With this, the Hamiltonian can be split into four terms 
\begin{equation}
\hat{H}=\hat{H}_{\rm mat}+\hat{H}_{\rm field}+\hat{H}_{\rm EP}+\hat{H}_{\rm PP},
\label{equ:H_tot}
\end{equation}
which will be discussed in more detail in the following.
    	
\paragraph{Material}
The isolated material consists of a collection of $N$ interacting dipoles, that are arranged on a two-dimensional square lattice with lattice parameter $a$. The Hamiltonian is given by
\begin{equation}
\label{matterhmat}
\hat{H}_{\rm mat}=\hat{H}_{0}+\hat{H}_{\rm nn},
\end{equation}
where $\hat{H}_{0}$ corresponds to the isolated dipoles, and $\hat{H}_{\rm nn}$ is a static nearest-neighbor interaction. We approximate the dipoles as simple two-level-systems. Hence, they can be represented by the Pauli-operators $\bm{\hat{\sigma}}_r=(\hat{\sigma}_r^1,\hat{\sigma}_r^2,\hat{\sigma}_r^3)^T$, where $\hat{\sigma}_r^3$ measures the difference in occupation of the two states at site $r$. The non-interacting Hamiltonian reads
\begin{equation}
\hat{H}_{0}=\frac{\Delta}{2}\sum_{r}\hat{\sigma}_r^3,
\end{equation}
with a level splitting $\Delta$, and the interaction reads
\begin{equation}
\label{hnn}
\hat{H}_{\rm nn}=-\frac{\alpha}{4}\sum_{\langle r,r'\rangle}\hat{\sigma}_r^1\hat{\sigma}_{r'}^1,
\end{equation}
where $\langle...\rangle$ indicates the sum over nearest neighbor pairs, and the parameter $\alpha$ controls the strength of the direct interaction. Moreover, we associate an in-plane dipole moment $\hat{\bm{p}}_r=p_0\hat{\sigma}_r^1\bm{e}_z$ to each two-level system. 
The corresponding transition matrix element $p_0$ will enter the light-matter coupling below.

The model can be viewed, e.g., as a two-level approximation for any kind of continuous model, where each site features one ion moving in an effective double-well potential. In that case, the two states represent the symmetric and anti-symmetric lowest lying energy states for the isolated site. A hybridization of the corresponding electronic wave functions gives rise to an asymmetric orbital and, thus, to a non-vanishing average electric dipole moment. For instance, the model can be considered as a minimal model for an ionic crystal, where the two states $\sigma_r^z=\pm1$ correspond to two positions of an ion within the unit cell, and the interaction arises both from a (partially screened) Coulomb interaction and from the configurational energy of the lattice distortion.

Below, we choose the parameters such that the model reproduces the dielectric properties of SrTiO$_3$ (see Sec.~IV). Of course, the model is highly oversimplified with respect  to a real material. In particular, the connection to SrTiO$_3$ is on a purely phenomenological level, and there is no direct correspondence between the parameters of our Hamiltonian and any microscopic quantities of the material. 
In particular, the connection to SrTiO$_3$ is on a purely phenomenological level, and there is no direct correspondence between the parameters of our Hamiltonian and any microscopic quantities of the material.
Nevertheless, our model should serve the purpose to (i), explain the application DMFT to light-matter coupled systems, and (ii), demonstrate that the present setting is a promising pathway to engineer the ferroelectric transition.	
Possible extensions of the model and the formalism will be briefly addressed in the Conclusions (Sec.~\ref{sec:conclusion}).

\paragraph{SPP mode}

For the description of the SPP mode, we start from the macroscopic theory of Ref.~\cite{Economou69} (see also Ref.~\cite{Ashida2020}), which treats the dielectric and the metal macroscopically, i.e., they enter via their dielectric function $\epsilon(\omega)$. Within this macroscopic approach, we determine the classical mode functions, then quantize the theory, and later integrate out the quantized modes to obtain the induced interaction within the 2D material. In principle, one could have started from a fully microscopic theory, in which the electronic degrees of freedom inside the metal are treated explicitly. However, in the end we are only interested in the effect of the SPP mode on the two-dimensional material, and this effect enters through a light-induced interaction. Due to the linearity of the description outside the 2D material, the induced interaction which would be obtained by (A), integrating out both the metal and field with a linear coupling of the currents in the metal and the field, and (B), the approach taken here to quantize the classical mode functions as described above, and then integrating out the modes, are equivalent.

 Due to the geometry  of the system (in-plane transition dipole moments) only the transverse magnetic (TM) SPP-mode couples to the material. For this mode, the transverse electric field can be expanded in the bosonic annihilation/creation operators $\hat{a}_{\bm q}$/ $\hat{a}_{\bm q}^\dagger$ as
\begin{equation}
\hat{\bm E}(\bm r)=\sum_{\bm q}\sqrt{\frac{\omega_{\bm q}}{2\epsilon_0\epsilon(x,\omega_{\bm q})Na^3}}\left[\bm{u}_{\bm q}(x)e^{i\bm{q}\cdot\bm{\rho}}\hat{a}_{\bm q}+h.c\right],
\label{equ:displacement_field}
\end{equation}
where $\bm{q}=(q_y,q_z)^T$ is a two-dimensional wave vector, $\bm{\rho}=(y,z)^T$ gives the position in the $y$-$z$-plane and $\omega_{\bm q}$ denotes the dispersion relation of the SPP mode.  The mode functions $\bm{u}_{\bm q}(x)e^{i\bm{q}\cdot\bm{\rho}}$ form an orthogonal basis and satisfy the transversality condition $\nabla\cdot\bm{u}_{\bm q}(\bm r)e^{i\bm{q}\cdot\bm{\rho}}=0$. They define the spatial structure of the mode. A detailed derivation, and the expressions for 
$\bm{u}_{\bm q}(x)$
can be found in App.~\ref{sec:quant_SPP}. Most importantly, the mode functions decay exponentially  like   $e^{Q_{\rm  m}x}$  and $e^{-Q_{\rm  d}x}$ in the metallic and dielectric region,  respectively, with real decay constants that asymptotically approach the value $Q_{\rm m}, Q_{\rm d} \sim |q|$ for large $|q|$. The latter implies that the distance $x_0$ between the material and interface controls the range of SPP momenta which couple to the material.

\begin{figure}
\centering
\includegraphics[width=\linewidth]{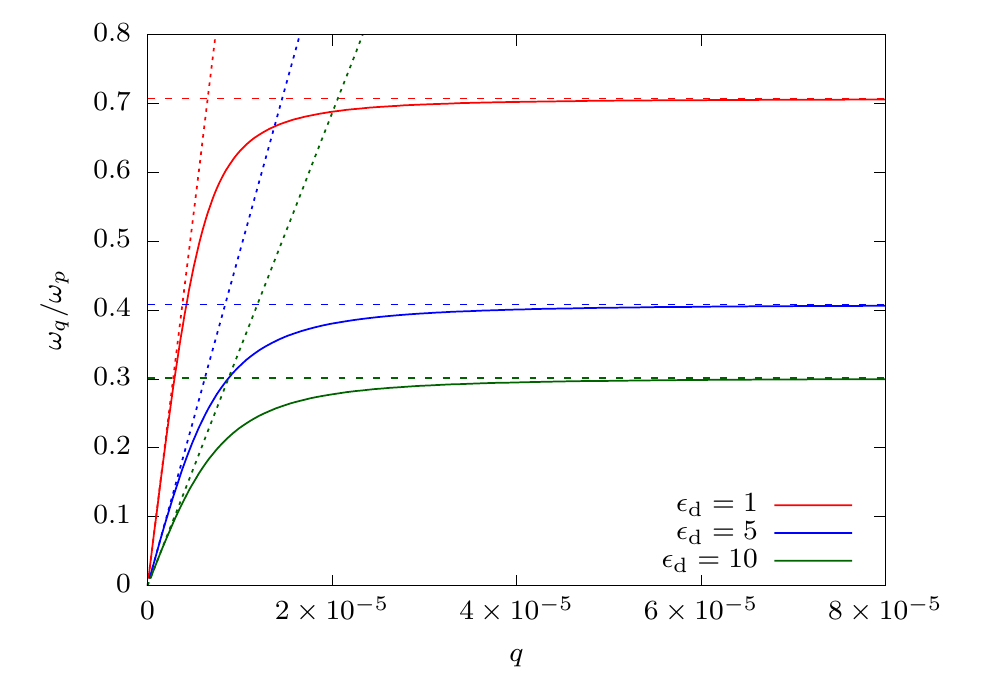}
\caption{Dispersion relation of the SPP-mode for three different values of the dielectric permittivity $\epsilon_{\mathrm{d}}$ and a Drude response of the metal. For small $q$, the dispersion increases linearly with slope $c/\sqrt{\epsilon_{\mathrm{d}}}$ (dotted lines). For $q\rightarrow\infty$, it approaches the constant value $\omega_{\rm p}/\sqrt{1+\epsilon_{\mathrm{d}}}$ (dashed lines). 
The unit of length has been set to the lattice parameter $a=3.9$\AA\ 
which is used for the model below.
}
\label{fig:disp}
\end{figure}

The corresponding free field Hamiltonian takes the form
\begin{equation}
\hat{H}_{\rm field}= \sum_{\bm q} \omega_{\bm q}\hat{a}_{\bm q}^\dagger\hat{a}_{\bm q},
\end{equation}
where $\omega_{\bm q}$ is the SPP dispersion. In the following, we assume that the electric permittivity in the metallic region is given by the simple Drude response
\begin{equation}
\epsilon_{\rm m}(\omega)=1-\left(\frac{\omega_{\rm p}}{\omega}\right)^2,
\end{equation}
where $\omega_{\rm p}$ denotes the plasma frequency. In this case, the dispersion relation $\omega_{\bm q}$ starts with a linear slope of $c/\sqrt{\epsilon_{\mathrm{d}}}$ at small $q$ and approaches a constant value of $\omega_{\rm p}/\sqrt{1+\epsilon_{\mathrm{d}}}$ for $q\rightarrow\infty$. This is displayed in Fig.~\ref{fig:disp} for three different values of $\epsilon_{\mathrm{d}}$. In our calculations, we set $\epsilon_{\rm d}=1$ in the dielectric region. 
With this,
the dispersion relation can be solved analytically for the frequency and reads
\begin{equation}
\omega_{\bm q}=\sqrt{\frac{\omega_{\rm p}^2}{2}+\bm{q}^2c^2-\sqrt{\frac{\omega_{\rm p}^4}{4}+\bm{q}^4c^4}}.
\end{equation}
	
\paragraph{Light-matter coupling} The interaction between light and matter is formulated in the dipolar gauge. This representation can be obtained from the minimal coupling Hamiltonian by a multi-center Power-Zienau-Woolley transformation \cite{Li2020} and gives rise to two additional terms in the Hamiltonian. The first one describes a linear coupling between the electric diplacement field $\hat{\bm E}(\bm r)$ and the polarization  $\hat{\bm{p}}_r=p_0\hat{\sigma}_r^1\bm{e}_z$ of the emitters and reads 
\begin{equation}
\hat{H}_{\rm EP}= \sum_{r,\bm q}\sqrt{\frac{\omega_{\bm q}}{2N}}\left[g_{\bm q}e^{i\bm{q}\cdot\bm{R}_{r}}\hat{a}_{\bm q}+h.c.\right]\hat{\sigma}_r^1,
\end{equation}
where the coupling constants are given by 
\begin{equation}
g_{\bm q}=\frac{1}{\sqrt{\epsilon_0\epsilon_{\rm d}a^3}}\bm{u}_{\bm q}(x_0)\cdot p_0\bm{e}_z,
\end{equation}
and $\bm{R}_r$ 
is the two-dimensional lattice vector of site $r$, and thus represents the ${\bm \rho}$-vector at which the electric field \eqref{equ:displacement_field} is evaluated. 
The second one is a non-linear term of the form
\begin{equation}
\hat{H}_{\rm PP}=\sum_{r,r'}\sum_{\bm q}\frac{|g_{\bm q}|^2}{2N}e^{-i\bm q\cdot(\bm R_r-\bm R_{r'})}\hat{\sigma}_r^1\hat{\sigma}_{r'}^1,
\label{hppp}
\end{equation}
which ensures that the total Hamiltonian is positive definite. 
The positive definite form of the Hamiltonian can be verified by re-writing the light-matter part and the free field term as $\hat{H}_{\rm field}+\hat{H}_{\rm EP}+\hat{H}_{\rm PP}=\sum_{\bm q}\omega_{\bm{q}}\hat{b}_{q}^{\dagger}\hat{b}_{q}$ with shifted field operators $\hat{b}_{\bm q}=\hat{a}_{\bm q}+\pi_{\bm q}$, where $\pi_{\bm q}=\sum_{r} \frac{g_{\bm q}}{\sqrt{\omega_{\bm{q}}2N}}e^{i{\bm q}\cdot{\bm R}_r}\hat{\sigma}_r^1$ is related to the polarization field. Note that $\hat{H}_{\rm PP}$ does not depend on the photon operators but still vanishes if the light-matter coupling is set to zero.
	
\subsection{Imaginary time action}
\label{sec:eff_action}

In the next step, we derive an effective description for the material that only depends on the matter degrees of freedom. Only the most important points are given in the main text, leaving details to App.~\ref{sec:ind_int_app}. We switch to an imaginary-time path integral formalism, where the model can be described in terms of an action
\begin{equation}
\label{action01}
S=S_{\rm{mat}}+S_{\rm{PP}}+S_{\rm{EP}}+S_{\rm{field}},
\end{equation}
which has been split into four terms analogous to the total Hamiltonian \eqref{equ:H_tot}.
Tracing out the photon fields yields an effective action of the form
\begin{equation}
S_{\rm eff}=S_{\rm mat}+S_{\rm ind},
\end{equation}
where the effect of the light-matter interaction is contained in an induced term $S_{\rm ind}$ that is defied by the relation
\begin{equation}
e^{-S_{\rm{ind}}=}e^{-S_{\rm{PP}}}\int\mathcal{D}[\bar{a},a]e^{-(S_{\rm{EP}}+S_{\rm{field}})},
\end{equation}
and 
takes the form of a retarded dipole-dipole interaction 
\begin{equation}
\label{sind}
S_{\rm{ind}}=-\frac{1}{2}\int\limits_0^\beta d\tau
\,
d\tau'\sum_{r,r'}\sigma_r^1(\tau)W^{\rm ind}_{r,r'}(\tau-\tau')\sigma_{r'}^1(\tau').
\end{equation}
The Matsubara and $\bm k$-space representation of the interaction vertex is given by
\begin{equation}
\label{wking}
W_{\bm k}^{\rm ind}(i\nu_n)=-|g_{\bm k}|^2 + |g_{\bm k}|^2\frac{\omega_{\bm k}^2}{\nu_n^2+\omega_{\bm k}^2},
\end{equation}
where the first term originates from the direct interaction \eqref{hppp}, and the second from integrating out the SPP mode. Using the replacement $i\nu_n\rightarrow\omega+i0$, this expression can be analytically continued to real frequencies. Interestingly, 
the interaction
vanishes at $\omega=0$; therefore, the effect of the light-matter interaction cannot be captured within 
the static
mean-field approximation (see Sec.~\ref{sec:results}). 
It is important to note that the cancellation of the two terms in two contributions in Eq.~\eqref{wking} at $\omega=0$ requires  $\hat{H}_{\rm EP}$ and $\hat{H}_{\rm PP}$ to be consistent, so that the positive definiteness of the Hamiltonian is preserved.  A simple understanding of the cancellation is that for the positive definite Hamiltonian a static shift of the polarization field can be absorbed in a shift of the photon operators, as explained below Eq.~\eqref{hppp}, and can therefore not give any contribution to the action. 
	
\section{Dynamical mean-field theory}
\label{sec:BDMFT}
In the last decades, DMFT has become a powerful tool to study the properties of quantum many-body systems. The main idea of this technique is to map a lattice model to an effective impurity problem. This is done by focusing on a single site
of the lattice
and incorporating the interaction with all other sites in the parameters of an effective environment. If the self-energy is assumed to be local in space, these parameters can be related to the corresponding lattice quantities by a set of self-consistent equations \cite{Georges1996}.  In this section we apply the DMFT formalism to the system under study. In particular, we introduce bosonic degrees of freedom and, therefore, follow a bosonic DMFT approach.

\subsection{Bosonic representation of the model}
\label{sec:HS-trafo}

As shown in Sec.~\ref{sec:eff_action}, the SPP mode mediates an effective interaction between the individual emitters. Similar to the static nearest-neighbor interaction, it couples the dipolar moments on different sites. We therefore start from the imaginary-time formalism, and represent the effective action for the matter as $S_{\rm eff}=S_{0}+S_{\rm int}$, 
where $S_0$ corresponds to the non-interacting dipoles and $S_{\rm int}$ is a retarded dipole-dipole interaction that combines the photon-mediated interaction \eqref{sind} and the contribution from the direct interaction Hamiltonian \eqref{hnn}, 
\begin{equation}
S_{\rm{int}}=-\frac{1}{2}\int\limits_0^\beta d\tau
\,
d\tau'\sum_{r,r'}\sigma_r^1(\tau)W_{r,r'}(\tau-\tau')\sigma_{r'}^1(\tau').
\end{equation}
The combined interaction vertex reads
\begin{equation}
W_{\bm k}(i\nu_n)=\frac{\alpha}{2}\left[\cos(k_y)+\cos(k_z)\right]
-|g_{\bm k}|^2\frac{\nu_n^2}{\nu_n^2+\omega_{\bm k}^2},
\end{equation}
where the first term is the static nearest neighbor interaction \eqref{hnn} in momentum space, and the second term derives from Eq.~\eqref{wking}.

Our goal is to apply bosonic DMFT to study effects beyond the mean-field limit. For that purpose, we perform a Hubbard-Stratonovich transformation, which introduces bosonic auxiliary fields $\varphi_r(\tau)$ at each site and define a new action $S_{\rm HS}$ such that the original action is reproduced if the auxiliary fields are traced out, i.e.,
\begin{equation}
e^{-S_{\rm eff}}=\int\mathcal{D}[\varphi]e^{-S_{\rm HS}}.
\end{equation}
Up to an irrelevant constant 
which will be
omitted here and in the following, the Hubbard-Stratonovich action reads
\begin{equation}
S_{\rm HS}=S_0+S_{\varphi\varphi}+S_{\varphi\sigma},
\label{equ:S_HS}
\end{equation}
with a quadratic term
\begin{equation}
S_{\varphi\varphi}=	\frac{1}{2}\int\limits_0^{\beta}d\tau
\,
d\tau'\sum_{r,r'}\varphi_r(\tau)[W^{-1}]_{r,r'}(\tau-\tau')\varphi_{r'}(\tau'),
\end{equation}
and a local linear coupling term
\begin{equation}
S_{\varphi\sigma}=-\sum_{r}\int\limits_0^{\beta}d\tau\varphi_r(\tau)\sigma_r^1(\tau).
\end{equation}

\subsection{Impurity action and self-consistent equations}

The Hubbard-Stratonovich action essentially defines a bosonic field theory on the lattice, with an anharmonic self-interaction that is introduced via the coupling of the field $\varphi_r$ and the spin $\sigma_r$. The local nature of this anharmonic interaction allows to follow a bosonic DMFT approach, and map the Hubbard-Stratonovic action \eqref{equ:S_HS} to a local impurity problem.
The mapping
can be achieved using the so-called cavity method \cite{Georges96,Ayral2013}:
One focuses  on a single site $r=c$, and integrates out all degrees of freedom related to  other sites of the lattice. 
Then a cumulant expansion up to second order is performed (see App.~\ref{sec:map_to_imp}), which yields an impurity action with quadratic and linear terms in the field $\varphi_c(\tau)$
\begin{align}
&S_{\rm HS}^{\rm imp}=S_{0}^c+S_{\varphi\varphi}^{\rm imp}+S_{\varphi\sigma}^{\rm imp}
\label{equ:imp_HS1},
\\
&S_{\varphi\sigma}^{\rm imp} =-\int\limits_0^{\beta}d\tau\varphi_c(\tau)[\sigma^1_c(\tau)-h(\tau)] ,
\\
&S_{\varphi\varphi}^{\rm imp} =\frac{1}{2}\int\limits_0^{\beta}d\tau
\,
d\tau'\varphi_{c}(\tau)\mathcal{W}^{-1}(\tau-\tau')\varphi_c(\tau').
\end{align}
Here $S_{0}^c$ is the action of a single isolated emitter. As in DMFT for fermionic systems, the interaction of the impurity with the rest of the lattice has been incorporated in an effective Weiss field $\mathcal{W}$. Moreover, there is an additional static field $h$ that couples to the auxiliary field $\varphi_c(\tau)$. It accounts for the fact that $\varphi_c(\tau)$ is a bosonic field and, therefore, may acquire a finite expectation value \cite{Anders2011}.
	
From the impurity action, one can calculate the local expectation value $\langle\varphi_c(\tau)\rangle_{S_{\rm HS}^{\rm imp}}$ and the local connected correlation function
\begin{equation}
U_{c,c}(\tau)=\langle \mathcal{T}\varphi_c(\tau)\varphi_c(0)\rangle_{S_{\rm HS}^{\rm imp}}^{\rm con}.
\end{equation} This defines the self-energy on the impurity via the Dyson equation
\begin{equation}
\Pi_{\rm loc}=\mathcal{W}^{-1}-U_{c,c}^{-1}.
\label{equ:self-energy1}
\end{equation}
On the other hand, the real space lattice Green's function
\begin{equation}
U_{r,r'}(\tau)=\langle \mathcal{T} \varphi_r(\tau)\varphi_{r'}(0)\rangle_{S_{\rm HS}}^{\rm con}
\end{equation}
is given by the Dyson equation in momentum space
\begin{equation}
U_k=W_k[1-\Pi_{k} W_k]^{-1}
\end{equation}
with the $k$-dependent self-energy $\Pi_{k}$. Within DMFT, the latter is assumed to be purely local and can be replaced by the impurity self-energy $\Pi_{\rm loc}$. Thus, by summing over the momentum dependent Green's function one obtains
\begin{equation}
U_{c,c}=\frac{1}{N}\sum_{\bm k}W_k[1-\Pi_{\rm loc} W_k]^{-1}
\label{equ:U_cc}
\end{equation}
for the local correlation function. 
This result can be used to determine the Weiss field $\mathcal{W}$ from Eq.~\eqref{equ:self-energy1}. The latter must be consistent with the one introduced in the impurity model. 
Moreover, imposing the condition $\langle\varphi_r(\tau)\rangle_{S_{\rm HS}}=\phi$ for all $r$, the external field is given by
\begin{equation}
h=\left[W_{mf}^{-1}-\mathcal{W}^{-1}_0\right]\phi,
\end{equation}
where
\begin{equation}
\mathcal{W}^{-1}_0=\mathcal{W}^{-1}(i\nu_{n=0}).
\end{equation}
These expressions form a closed set of equations. 
Further details on the 
derivation can be found in App.~\ref{sec:deriv_sc}.

\subsection{Impurity model}

The impurity action \eqref{equ:imp_HS1} defines a generalized spin boson model, which couples a two-level system to a continuum of modes with propagator $\mathcal{W}$. To solve the impurity model, it is more convenient to integrate out the Hubbard-Stratonovich fields, leading to the action of a spin with retarded interactions.	The new impurity action reads (see  App.~\ref{sec:elim_aux_fields})
	\begin{equation}
		S^{\rm imp}=S^{c}_{0}+S_{{\rm int},1}^{\rm imp}+S_{{\rm int},2}^{\rm imp}
		\label{equ:S_imp}
	\end{equation}
	with the linear interaction term 
	\begin{equation}
		S_{{\rm int},1}^{\rm imp}=b\int\limits_0^\beta d\tau \sigma^1_c(\tau),
	\end{equation}
	and the retarded interaction term
	\begin{equation}
		S_{{\rm int},2}^{\rm imp}=-\frac{1}{2}\int\limits_0^\beta d\tau\int\limits_0^\beta d\tau'\,\sigma^1_c(\tau)\mathcal{W}(\tau-\tau')\sigma^1_c(\tau').
	\end{equation}
For the solution of this model we use the strong coupling expansion introduced in Refs.~\cite{Aaram2022,Aaram2021}, which is based on a summation of the skeleton expansion of the partition function in terms of the retarded propagator. We remark that the present problem could also be addressed by a standard hybridization expansion \cite{Werner2006}, but even within the diagrammatic approach the perturbation order can be increased to convergence, so that the results can be considered as 
numerically 
exact. 
With the connected correlation function
\begin{equation}
\chi_{c,c}(\tau)=\langle \mathcal{T} \sigma_c^1(\tau)\sigma_c^1(0)\rangle_{S^{\rm imp}}^{\rm con},
\end{equation}
the Green's function for $\varphi_c(\tau)$ is given by
\begin{equation}
U_{c,c}=\mathcal{W}+\mathcal{W}\chi_{c,c}\mathcal{W}.
\end{equation}
Using Eq.~\eqref{equ:self-energy1}, this yields the following expression for the self-energy 
\begin{equation}
\Pi_{\rm loc}=\left[1+\chi_{c,c}\mathcal{W}\right]^{-1}\chi_{c,c}.
\label{equ:self-energy}
\end{equation}
Likewise, the external field $b=\mathcal{W}_0h$ can be written in terms of $\langle\sigma^1\rangle$ and reads
\begin{equation}
b=\left[\mathcal{W}_0-W_{mf}\right]\langle\sigma^1\rangle,
\label{equ:b}
\end{equation}
which closes the self-consistency.
	
	\subsection{Numerical implementation}
	\label{sec:numerics}
	
\begin{figure}
\centering
\includegraphics[width=\linewidth]{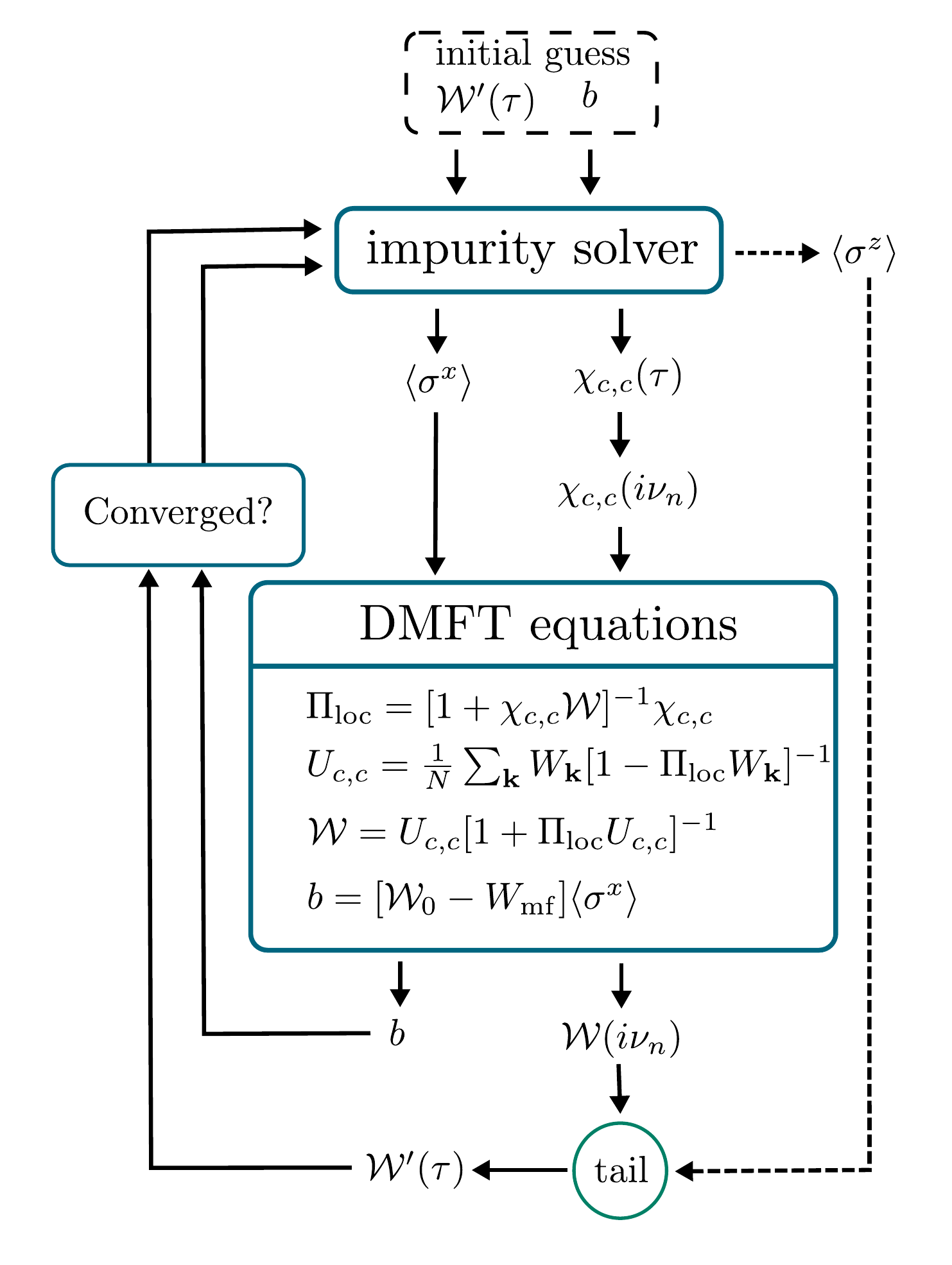}
\caption{Illustration of the DMFT loop.}
\label{fig:algo}
\end{figure}

The self-consistent DMFT equations are solved in an iterative procedure. 
The basic algorithm, which is illustrated in Fig.~\ref{fig:algo}, consists of two major components. 
On the one hand, there is the impurity solver that allows to calculate local expectation values and correlation functions on the impurity. It is based on a strong coupling expansion in the imaginary-time domain similar to the one discussed in Refs.~\cite{Aaram2022,Aaram2021}.  However, there is an additional subtlety regarding the Weiss field. In general, 
the latter
may contain an instantaneous contribution proportional to $\delta(\tau)$ that corresponds to a frequency independent part in the Matsubara representation; thus, we write $\mathcal{W}(\tau)=\mathcal{W}'(\tau)+w_0\delta(\tau)$, where $\mathcal{W}'(\tau)$ does not contain an instantaneous part. However, since $(\hat{\sigma}_c^1)^2=\mathbb{I}$, the part proportional to $\delta(\tau)$ only gives rise to a constant shift of the energy and does not influence the values of the correlation functions. Therefore, we can omit 
the instantaneous contribution
and only use $\mathcal{W'(\tau)}$ as an input.
	
On the other hand, there is the set of self-consistent equations that allow to calculate $\mathcal{W}$ and $b$ from $\chi_{c,c}$ and $\langle\sigma^1\rangle$. This is done in Matsubara space, since all quantities are diagonal in this representation. 
Therefore, we need to perform a Fourier transform when passing from the impurity solver to the DMFT equations and vice versa.  
Some
extra care has to be taken for the inverse transform from the Matsubara representation to the imaginary-time domain,
do deal with the numerical cutoff in the frequency summations. We
use an analytical estimate for the high-frequency tail of the Weiss field which is given by 
\begin{equation}
\mathcal{W}(i\nu_n)	\sim w_0+\frac{w_2}{(i\nu_n)^2},
\end{equation}
where the constants $w_0$ and $w_2$ 
can be expressed in terms of 
the light-matter coupling strength, the parameters of the material and the expectation value $\langle\sigma^z\rangle$. The corresponding equations and a detailed derivation are given in App.~\ref{sec:tail_app}. Moreover, we define a function
\begin{equation}
f(i\nu_n)=w_0+\frac{w_2}{(i\nu_n)^2-\delta^2}
\end{equation}
 that shows the same behavior at large $i\nu_n$, where $\delta\ll 1$ is a small regulator. Then, the difference $\mathcal{W}(i\nu_n)-f(i\nu_n)$ 
decays at least as $1/\nu_n^4$ for large frequencies, 
such that even the numerical Fourier transform yields a sufficiently smooth result. Finally, we may add the imaginary time representation of $f$, which is analytically known. Since the impurity solver only requires the non-instantaneous part $\mathcal{W}'(\tau)$, we already exclude the instantaneous part
 from the analytical Fourier transform of $f$, i.e.
\begin{equation}
\mathcal{W}'(\tau)=\mathcal{F}^{-1}\left\{\mathcal{W}(i\nu_n)-f(i\nu_n)\right\}+f'(\tau),
\end{equation}
where $f'(\tau)$ denotes the analytical inverse Fourier transform of $f(i\nu_n)-w_0$ and is given by
\begin{equation}
f'(\tau)=\frac{w_2}{2\delta}\frac{\cosh[\delta(\tau-\beta/2)]}{\sinh[\delta\beta/2]}.
\end{equation}

In summary, the DMFT loop consists of the following steps:
\begin{enumerate}
\item Start from an initial guess for $\mathcal{W'}(\tau)$ and $b$, and pass it to the impurity solver.
\item Calculate $\chi_{c,c}(\tau)$, $\langle\sigma^1\rangle$ and $\langle\sigma^z\rangle$.
\item Transform $\chi_{c,c}(\tau)$ to Matsubara space.
\item Insert $\chi_{c,c}(i\nu_n)$ and $\langle\sigma^1\rangle$ in the DMFT equations and calculate the new Weiss field $\mathcal{W}(i\nu_n)$ and the new value of $b$.
\item Apply the inverse Fourier transform,
with 
the tail correction to $\mathcal{W}(i\nu_n)$ using $\langle\sigma^z\rangle$, to obtain $\mathcal{W}'(\tau)$ as described above.
\item Check whether $\mathcal{W}(\tau')$ has converged. 
If not
pass the new $\mathcal{W}'(\tau)$ and $b$ to the impurity solver and repeat the procedure starting from step 2.
\end{enumerate}	
	
\section{Results}
\label{sec:results}

\subsection{Mean-field approximation}
\label{sec:MF}

\begin{figure}
\centering
\includegraphics[width=\linewidth]{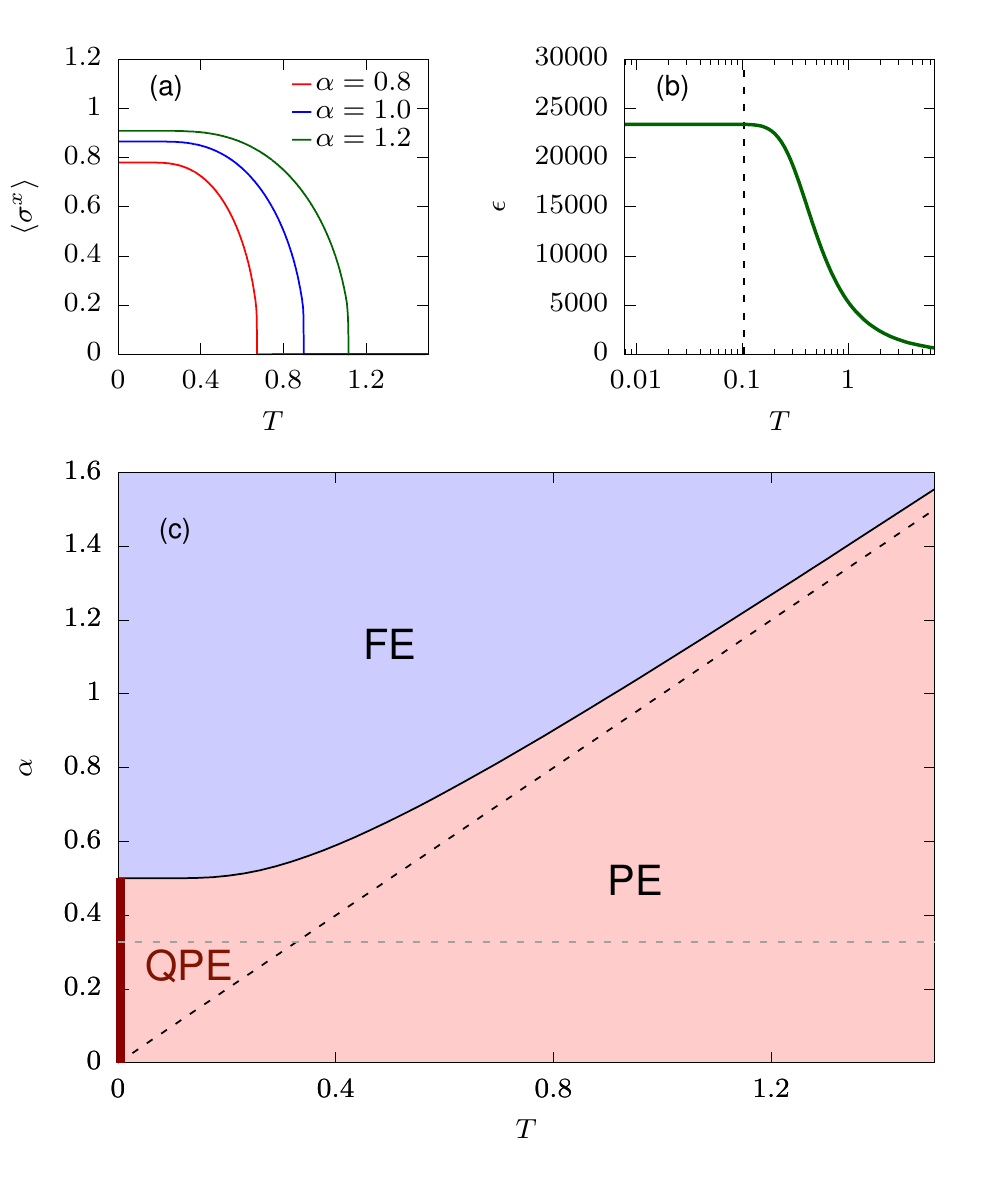}
\caption{(a) Order parameter, (b) dielectric constant and (c) phase diagram in the mean-field approximation. The unit of energy is the level splitting $\Delta$. 
The solid line in panel (c) indicates the boundary between the ferroelectric (blue shaded area) and the paraelectric (red shaded area) phase. The black dashed line shows the phase boundary $T=\alpha$ for the classical model, with $\Delta\to0$.  The quantum paraelectric state at low temperatures is marked by the dark red line. Panel (b) shows the temperature-dependent dielectric function $\epsilon(T)$ for a selected parameter $\alpha=0.328$ (gray dashed line in (c))  in this regime; $\epsilon(T)$ saturates to a constant value for sufficiently low temperature (see main text for the choice of $\alpha$).
}
\label{fig:mf}
\end{figure}

As emphasized in the introduction, a mean-field solution of the model cannot capture the effect of the transverse field on the phase transition.  The latter is entirely driven by the static nearest-neighbor interaction $\hat{H}_{\rm nn}$ within the mean-field approach.  Nevertheless, it is illustrating to discuss the mean-field solution of the model, as it gives the overall structure of the phase diagram without coupling to the SPP mode. The derivation of the mean-field equations is standard and can be found in App.~\ref{sec:MF_app}.  In the mean-field approximation, all spatial and temporal fluctuations are neglected, and we introduce a uniform time-independent order parameter $\langle\sigma^1\rangle\equiv\langle\sigma_r^1(\tau)\rangle\,$ for all $ r$ and $ \tau$. For $\langle\sigma^1\rangle=0$, the system is in the disordered paraelectric (PE) state, where the average polarization vanishes,  while for $\langle\sigma^1\rangle>0$, the material undergoes a transition to the ordered ferroelectric (FE) state with a nonzero electric dipole moment. The order parameter at inverse temperature $\beta$ is determined by  the self-consistent equation 
\begin{equation}
\langle\sigma^1\rangle= \frac{h_{\rm mf} \tanh\left(\beta\sqrt{(\Delta/2)^2+h_{\rm mf}^2}\right)}{\sqrt{(\Delta/2)^2+h_{\rm mf}^2}},
\label{equ:sc_equ_main}
\end{equation}
which is simply the expectation value of $\sigma^1$ for a two level system in a self-consistent field $h_{\rm mf}=\alpha\langle\sigma^1\rangle$. 
The PE solution $\langle\sigma^1\rangle=0$ to Eq.~\eqref{equ:sc_equ_main}
becomes unstable below a critical temperature, where $\langle\sigma^1\rangle$ may also take a non-vanishing value. Fig.~\ref{fig:mf}(a) shows the order parameter as a function of temperature for three different values of $\alpha$. It can be seen that $\langle\sigma^1\rangle$ saturates to a constant value at low temperatures and continuously drops to zero at some critical temperature $T_c$, i.e., the system undergoes a second order phase transition. The critical temperature as well as the overall strength of the order parameter increase as $\alpha$ is enhanced. In Fig.~\ref{fig:mf}(c), we show a phase diagram in the $T$-$\alpha$-plane. Due to the finite level splitting $\Delta$ of the emitters, the system exhibits a quantum paraelectric 
(QPE) regime for  $\alpha<\Delta/2$, where it  remains disordered down to zero temperature. 
In the following we will see how these findings change if additional temporal fluctuations 
and the coupling to the SPP mode
are included within the DMFT formalism.
	
\subsection{Model parameters and light induced interactions}

To couple the SPP mode to the system, we have to fix the model parameters to reasonable values. To quantify the strength of the light-matter interaction, we introduce the collective coupling
\begin{equation}
\lambda=\frac{p_0^2}{\epsilon_{\rm d}\epsilon_0a^3},
\end{equation}
with the transition dipole moment $p_0$, and the lattice constant $a$ of the material. For all simulations below, we set $\lambda=4024\Delta$. This value has been extracted from a fit of the mean-field result for the dielectric constant to experimental data for the paradigmatic quantum paraelectric material SrTiO$_3$ \cite{Mueller1979}, which yields $\Delta=3.3$meV, $\alpha=0.328\Delta$, and $\lambda=4024\Delta$ (see Ref.~\cite{Lenk2022}). For reference, the dielectric function obtained within the mean-field solution, which is given by $\epsilon(T)=1+\lambda\frac{\chi_{\rm at}}{1-\alpha\chi_{\rm at}}$ with the static susceptibility $\chi_{\rm at}=2\tanh(\Delta/2T)/\Delta$ for a single isolated two-level system \cite{Lenk2022}, is shown in Fig.~\ref{fig:mf}(b). In the following, we will keep $\Delta$ as an energy unit, fix the value of $\lambda$ for all simulations, and vary the direct interaction strength $\alpha$. The latter controls the transition temperature without coupling to light; in SrTiO$_3$, this can be done, e.g., by adding strain. The length scale $a$ will be set to the lattice constant $a=3.9$\AA\ in SrTiO$_3$ and the plasma frequency equals $\omega_{\rm p}=\sqrt{2}\Delta$.

\begin{figure}
\centering
\includegraphics[width=\linewidth]{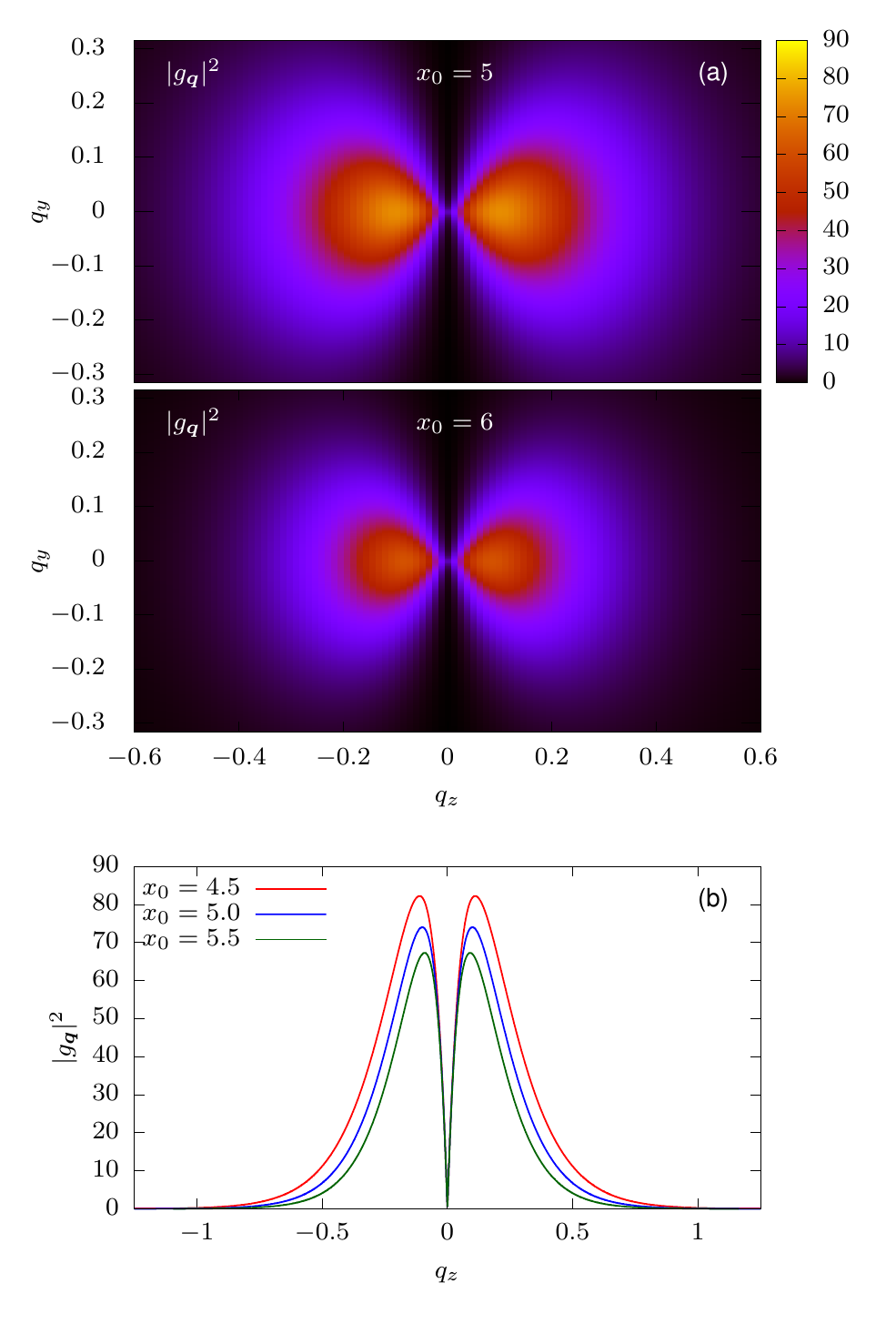}
\caption{
Momentum dependence of the light-matter coupling. (a) $|g_{\bm{q}}|^2$ at $x_0=5$ and $x_0=6$ for $\lambda=4024$. (b) $|g_{\bm{q}}|^2$ for the modes propagating along the $z$-axis and for three different values of $x_0$.  The unit of energy is given by 
$\Delta$, and the unit of length has been set to the lattice parameter $a$.}
\label{fig:gq2}
\end{figure}

With this, the $\bm{q}$-dependent coupling constants can be rewritten as $|g_{\bm{q}}|^2=\lambda (\bm{u}_{\bm q}(x_0)\cdot\bm{e}_z)^2$. In Fig.~\ref{fig:gq2}(a), the strength of the light-matter coupling for $x_0=5$ and $x_0=6$ is displayed as a function of the two-dimensional vector $\bm q$. It can be seen that only a small range of momenta contributes to the interaction. Comparing the plot for $x_0=5$ and $x_0=6$  illustrates how this interaction range can be controlled by the distance of the material from the interface, due to the exponential decay of the mode functions. This  provides the main pathway to engineer the ferroelectric transition in  the present setting. The decay of the interactions with distance $x_0$ and with $|q|$ is also seen from the line plots of the coupling strength $|g_{\bm{q}}|^2$ to modes propagating along the $q_z$-axis for three values of $x_0$ (Fig.~\ref{fig:gq2}(b)).

	\subsection{Paraelectric regime}

\begin{figure}
\centering
\includegraphics[width=\linewidth]{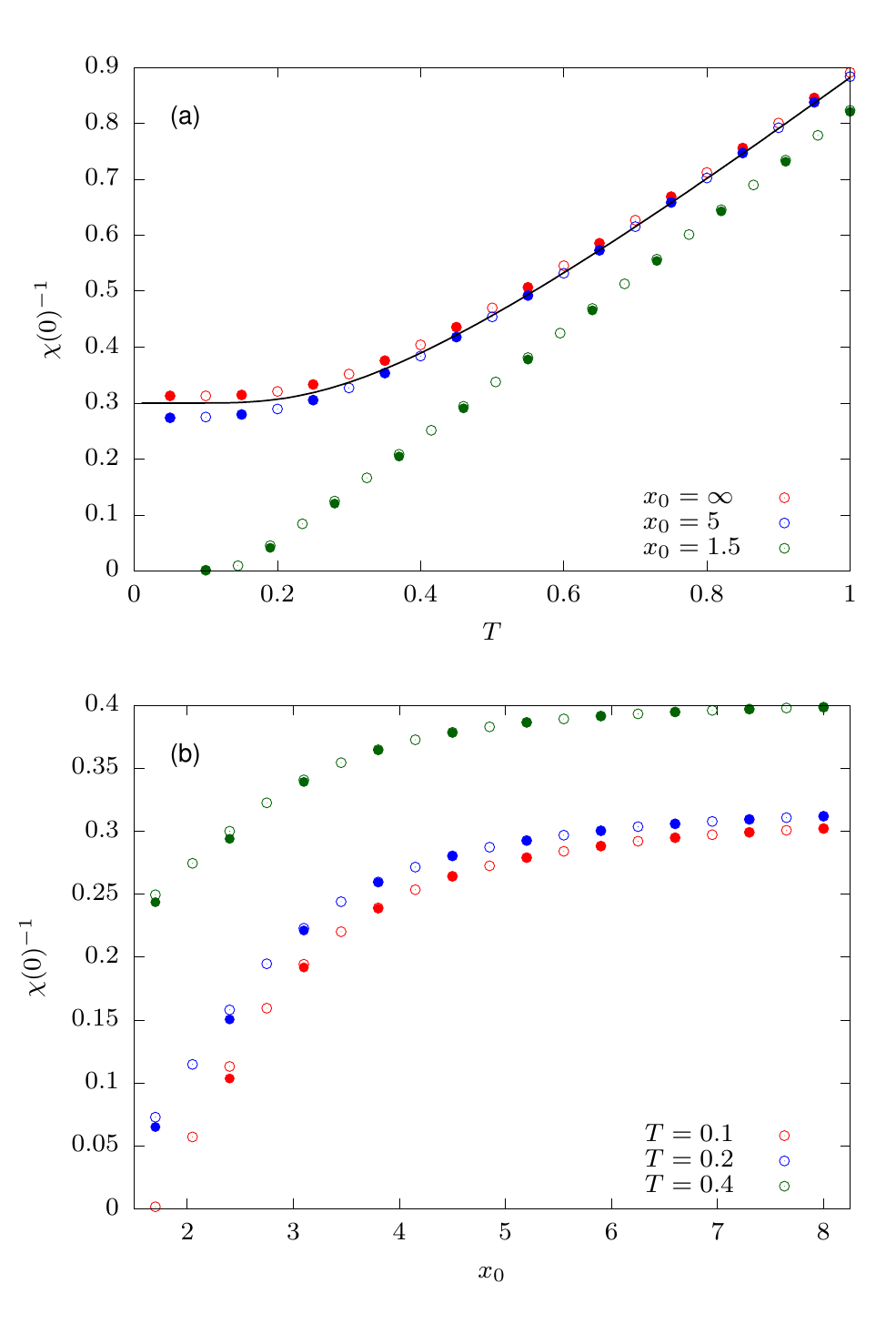}
\caption{
(a) Inverse susceptibility as a function of temperature for $\alpha=0.2$ at three different values of $x_0$. The black line indicates the corresponding mean-field result that does not depend on $x_0$. (b) Inverse susceptibility as a function of the distance $x_0$ for $\alpha=0.2$ at three different temperatures. In both panels, the system is in the PE regime. Empty and filled symbols correspond to the OCA and TCA solution of the DMFT impurity model, respectively (see text).
}
\label{fig:PE}
\end{figure}
	
In the PE regime, the average polarization of the solid vanishes. We therefore consider the static linear susceptibility $\chi(0)$ which measures the linear response of the average $\langle \sigma^1\rangle$ to an external field $B_{\rm ext}$. 
The susceptibility can be measured by adding the term $\hat{H}_{\rm ext}=-B_{\rm ext}\sum_{r}\hat{\sigma}_{r}^x$ to the matter Hamiltonian \eqref{matterhmat}, keeping it throughout the DMFT self-consistency 
(i.e., there is a term $S_{\rm ext}=-B_{\rm ext}\int_{0}^{\beta}d\tau\sigma_c^1(\tau)$ in the impurity action in addition to the self-consistent field Eq.~\eqref{equ:b}.) 
The susceptibility is calculated by the ratio $\chi(0)\approx\langle\sigma^1\rangle/B_{\rm ext}$, for fields $B_{\rm ext}$ which are sufficiently small that the response is in the linear regime.

Figure~\ref{fig:PE} shows the results for a nearest-neighbor interaction $\alpha=0.2$. This is a regime in which mean-field theory predicts a PE phase down to zero temperature (see Fig.~\ref{fig:mf}(c)). In Fig.~\ref{fig:PE}(a), the inverse susceptibility is plotted as a function of temperature for three different values of $x_0$. Data points marked with full and empty circles have been obtained for different orders of the diagrammatic impurity solver (one-crossing approximation (OCA) and two-crossing approximation (TCA Refs.~\cite{Aaram2022,Aaram2021} for more details). The two solvers yield consistent results, which shows that the results can be considered as converged.  Since the light-matter interaction is exponentially suppressed with increasing distance, there is no effect of the  SPP mode on the material at $x_0=\infty$.  One observes a lowering of the susceptibility for $x_0=\infty$ with respect to the static mean field result (dashed line), which is expected because  fluctuations beyond mean field (without coupling to the cavity) reduce the tendency towards symmetry breaking. However, if the material is moved closer to the metal surface, the coupling to the SPP mode is enhanced, which leads to an increasing effect  on the equilibrium properties of the system. As can be seen in the plot, the static susceptibility is enhanced (i.e. the inverse static susceptibility is decreased) due to the light-matter interaction. 

Fig.~\ref{fig:PE}(b) shows the inverse static susceptibility as a function of the distance $x_0$ for three different temperatures. Again, it is clearly visible that the susceptibility, and therefore the tendency towards ferroelectric ordering is enhanced as the material approaches the metallic surface. The overall effect is most pronounced at low temperatures.  Moreover, it can be seen that the OCA and TCA data start to deviate more strongly at small distances. This shows that the impurity model, which  describes the single atom of the lattice in an effective medium, is driven towards strong coupling as the coupling to the SPP is increased,  so that higher orders in the diagrammatic expansion become more significant.

\subsection{Ferroelectric regime}

\begin{figure}
\centering
\includegraphics[width=\linewidth]{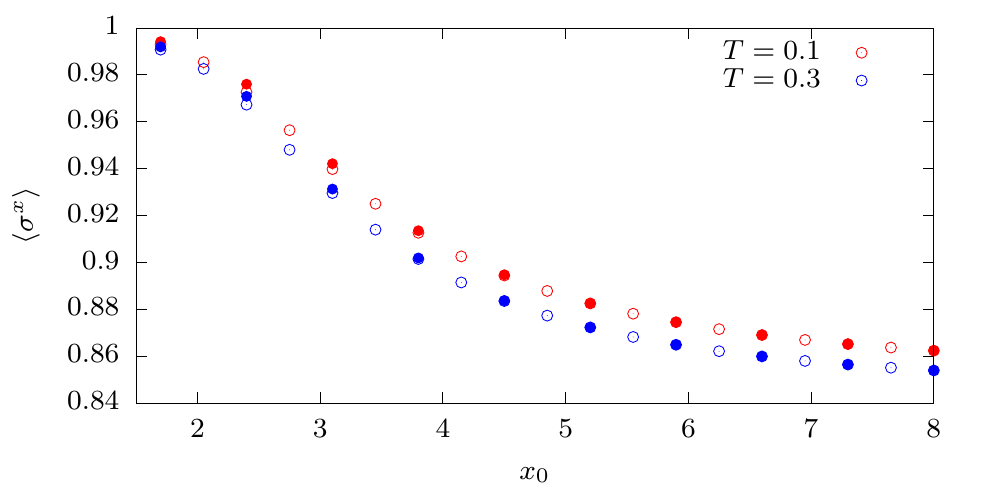}
\caption{Order parameter in the FE regime for $\alpha=1.0$ as a function of the distance $x_0$ at two different temperatures.  As in Fig.~\ref{fig:PE}, empty (filled) symbols correspond to data obtained with OCA (TCA).
}
\label{fig:FE}
\end{figure}
Below a critical temperature, the solid spontaneously acquires a non-vanishing average electric polarization and enters a FE state. The average polarization is proportional to the order parameter $\langle\sigma^1\rangle$. In Fig.~\ref{fig:FE}, we plot the order parameter for a system with  $\alpha=1$ as a function of $x_0$ for $T=0.1$ and $T=0.3$. For both temperatures, the order parameter $\langle\sigma^1\rangle$ increases as the distance is decreased, i.e., consistent with the tendency in the PE phase, the coupling to the SPP mode enhances the order in the solid.  Moreover, comparing the two curves shows that the increase is larger for $T=0.3$; thus, the effect becomes stronger if the system is closer to the phase transition.
	
\subsection{Phase diagram}
	
\begin{figure}
\centering
\includegraphics[width=\linewidth]{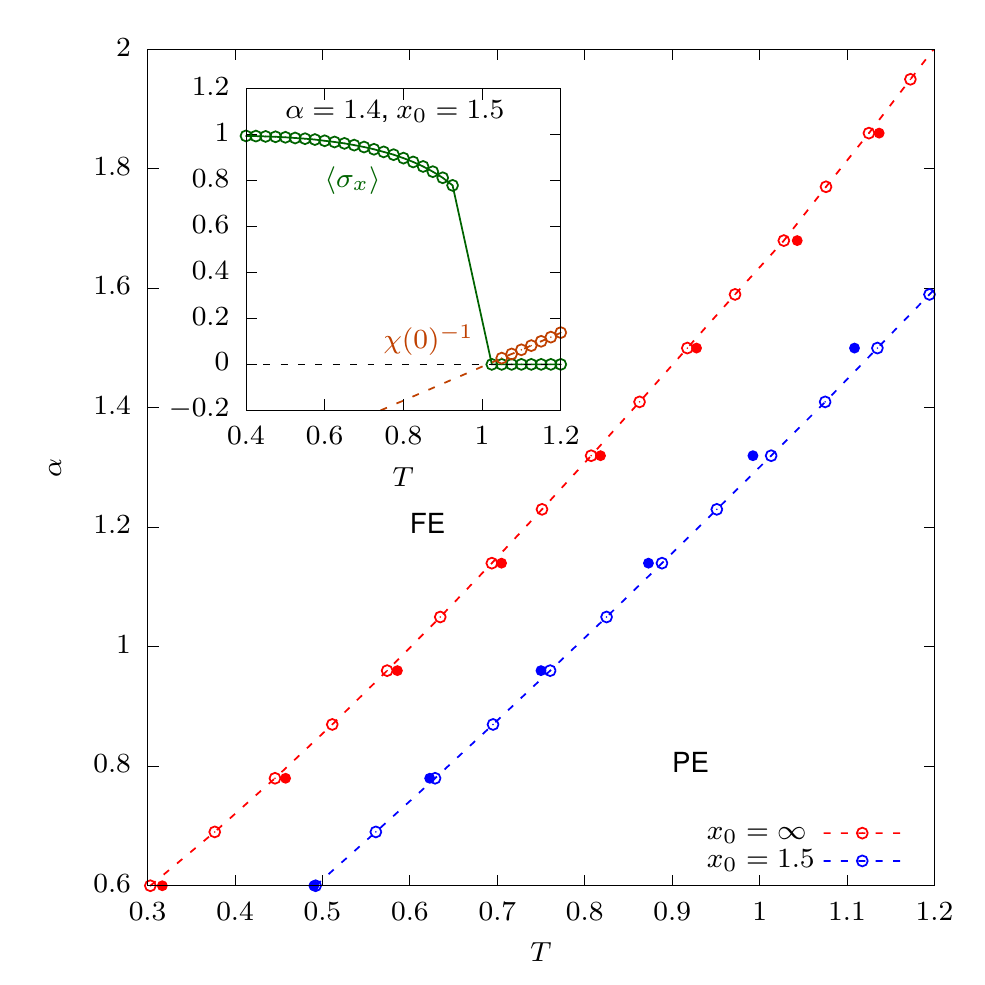}
\caption{Main panel: Phase diagram as a function of $\alpha$ and temperature. Empty (filled) symbols correspond to data obtained with OCA (TCA). Inset: Order parameter (green) and inverse susceptibility (orange) for $x_0=1.5$ at $\alpha=1.4$. The orange dashed line shows a linear fit to the data for the inverse susceptibility, from which the transition temperature $T_c$ is obtained.
}
\label{fig:phasediagram}
\end{figure}

Having discussed the effect of the SPP mode on the PE and the FE state, we  investigate how the phase diagram is influenced by the light-matter interaction. 
As it turns out, 
DMFT predicts the transition to be first order instead of second order. A similar behavior has been previously found in an application of DMFT to the standard lattice $\varphi^4$-theory \cite{Akerlund2013, Akerlund2014}. Nevertheless, apart from the first order nature of the phase transition,  DMFT has been shown in this case to provide a rather accurate description of the transition temperature and the properties of the system inside the ordered and disordered 
phase, compared 
to lattice Quantum Monte Carlo results.  In the present system, there exists a small coexistence region in which both the normal and the symmetry-broken phase can be stabilized. We take the divergence of the susceptibility as a measure for the lower bound of the coexistence region and the transition temperature $T_c$. This is illustrated in the inset of Fig.~\ref{fig:phasediagram}, where we plot the order parameter and the inverse susceptibility as a function of temperature for $x_0=1.5$ at $\alpha=1.4$. The order parameter exhibits a jump near $T=1$, which indicates the first order transition. Close to this point, the inverse static susceptibility $\chi(0)^{-1}$ crosses zero. 

The main panel of Fig.~\ref{fig:phasediagram} displays the transition temperature $T_c$ for $x_0=\infty$, i.e., without light-matter coupling, and for $x_0=1.5$. Again, the circles correspond to OCA results, whereas the dots represent data calculated with TCA. It can be seen clearly that the phase transition is shifted to higher temperatures/smaller $\alpha$ due to the coupling to the SPP mode, and as a consequence, the ferroelectric state is stabilized over a larger range of parameters. This is consistent with the increase of the susceptibility in the disordered phase, and the increase of the order parameter in the FE phase.
	
\section{Conclusion}
\label{sec:conclusion}

To summarize, we have shown that bosonic DMFT is a useful tool for the investigation of systems with light-matter coupling. It allows to account for light-induced effects that cannot be captured by a mean-field approximation due to the positive definite structure of the photon-mediated interaction. As an example, we have considered a simple model of a two-dimensional material that features a ferroelectric phase transition, which is driven  by a static nearest-neighbor dipole-dipole interaction. The material couples to a SPP mode supported by a metal-dielectric interface via its dipolar moments. We have seen that the coupling between light and matter leads to (i) an enhancement of the static electric susceptibility in the paraelectric regime and (ii) an increase of the order parameter in the ferroelectric phase. Both effects are most pronounced near the phase transition, and lead to a stabilization of the ferroelectric state over a larger range of parameters. 
Although the model is not a microscopic model for STO, its parameters can be chosen in an unbiased way to reproduce the linear macroscopic response of STO. Hence, from our results one may at least expect measurable effects on the phase transition temperature for settings similar to the ones considered in our work.

Intuitively, this behavior can be best explained from a macroscopic point of view: In general, the transition to the ferroelectric state is indicated by a softening of the corresponding collective mode at momentum $\bm q=0$. The linear mixing of this soft mode and the $\bm q=0$ SPP mode does not shift the phase transition (see discussion related to no go theorems). However, in the quantum paraelectric the soft mode is strongly renormalized by its anharmonic interaction with collective dipole fluctuations at all other momenta, which is in essence the reason for preventing the soft mode from condensation at the classical mean-field transition temperature. Now, the vacuum fluctuations of the SPP mode hybridize with  these collective dipole fluctuations at all scales, which can then, though the anharmonic mixing, affect the $\bm q=0$ soft mode at and near the phase transition.  This nonlinear mixing of nonzero frequency modes (which are affected by the cavity) and the static response of the material is similar to the phenomenological theory presented in Ref.~\cite{Ashida2020}. The results show that the coupling of  matter to surface plasmons near interfaces provides a promising pathway to engineer material properties with electromagnetic fields.

It would be interesting to see how well the DMFT approximation compares to other approaches 
to solve the 
present model.  However,  the microscopic description of the strongly correlated light matter problem is challenging even within the minimal model studied here:  Efficient lattice quantum Monte Carlo approaches for spin boson models have been developed \cite{Weber2022}, which may however suffer from a sign problem if the induced retarded interaction $W_{r,r'}(\tau)$ in the lattice model changes sign as a function of imaginary time. (Note that the vanishing of the interaction at zero frequency implies that $\int d\tau \, W_{r,r'}(\tau)=0$.) For one-dimensional systems on the other hand, matrix product state (MPS) algorithms would be a possibility. DMFT is expected to be most appropriate for high-dimensional systems, but the long-range nature of the light-induced interaction should be beneficial for the accuracy even in the case of low-dimensional system. For a comparison with MPS algorithms, however, the main challenge is the requirement to take into account a continuum of modes; this may be addressed in analogy to the MPS treatment of systems with electron phonon coupling \cite{Jeckelmann1999, Jansen2021}. 

A main advantage of DMFT is that the formalism can be readily adapted to more complex models, enabling in principle a  realistic description of strongly correlated electron-systems. In particular, this includes itinerant electrons and their interactions with the electromagnetic field, which could be treated using the extended DMFT formalism \cite{Sun2002} or diagrammatic extensions of DMFT similar to the GW+DMFT formalism \cite{Biermann2003,Ayral2013}. Moreover, in contrast to some methods commonly used in quantum optics, DMFT does not use the  Markov approximation when eliminating the photon propagator, but instead works with the fully frequency-dependent interaction for a continuum of  electromagnetic 
modes. It also includes 
both rotating and counter-rotating terms, which becomes particularly relevant in the regime of strong light-matter coupling. Finally, it should be remarked that the dielectric environment provided by the interface or cavity will not only affect the transverse electromagnetic field, but also the longitudinal field (direct Coulomb interaction). This will be relevant in particular for materials where the Coulomb interaction is weakly screened from the outset, and can therefore be further controlled by the dielectric environment \cite{Steinke2020}.  (In the present minimal model,  we have kept the direct interaction short-ranged, 
since the aim was to study the qualitative effect of the transverse field.)  As the extended DMFT formalism relies on a fully frequency dependent interaction, it should allow to consistently address the possibility of materials engineering by shaping both transverse and longitudinal fields at interfaces and in cavities.

\section*{Acknowledgments}
K.L and M. E. were funded by the ERC Starting Grant No. 716648, and by the Deutsche Forschungsgemeinschaft (DFG, German Research Foundation) – Project-ID 429529648 – TRR 306 QuCoLiMa (``Quantum Cooperativity of Light and Matter'').  J.L. is supported by SNSF Grant No. 200021-196966 and 
Marie Sklodowska Curie Grant Agreement No. 884104 (PSI-FELLOW-III-3i).


\begin{thebibliography}{58}%
		\makeatletter
		\providecommand \@ifxundefined [1]{%
			\@ifx{#1\undefined}
		}%
		\providecommand \@ifnum [1]{%
			\ifnum #1\expandafter \@firstoftwo
			\else \expandafter \@secondoftwo
			\fi
		}%
		\providecommand \@ifx [1]{%
			\ifx #1\expandafter \@firstoftwo
			\else \expandafter \@secondoftwo
			\fi
		}%
		\providecommand \natexlab [1]{#1}%
		\providecommand \enquote  [1]{``#1''}%
		\providecommand \bibnamefont  [1]{#1}%
		\providecommand \bibfnamefont [1]{#1}%
		\providecommand \citenamefont [1]{#1}%
		\providecommand \href@noop [0]{\@secondoftwo}%
		\providecommand \href [0]{\begingroup \@sanitize@url \@href}%
		\providecommand \@href[1]{\@@startlink{#1}\@@href}%
		\providecommand \@@href[1]{\endgroup#1\@@endlink}%
		\providecommand \@sanitize@url [0]{\catcode `\\12\catcode `\$12\catcode
			`\&12\catcode `\#12\catcode `\^12\catcode `\_12\catcode `\%12\relax}%
		\providecommand \@@startlink[1]{}%
		\providecommand \@@endlink[0]{}%
		\providecommand \url  [0]{\begingroup\@sanitize@url \@url }%
		\providecommand \@url [1]{\endgroup\@href {#1}{\urlprefix }}%
		\providecommand \urlprefix  [0]{URL }%
		\providecommand \Eprint [0]{\href }%
		\providecommand \doibase [0]{https://doi.org/}%
		\providecommand \selectlanguage [0]{\@gobble}%
		\providecommand \bibinfo  [0]{\@secondoftwo}%
		\providecommand \bibfield  [0]{\@secondoftwo}%
		\providecommand \translation [1]{[#1]}%
		\providecommand \BibitemOpen [0]{}%
		\providecommand \bibitemStop [0]{}%
		\providecommand \bibitemNoStop [0]{.\EOS\space}%
		\providecommand \EOS [0]{\spacefactor3000\relax}%
		\providecommand \BibitemShut  [1]{\csname bibitem#1\endcsname}%
		\let\auto@bib@innerbib\@empty
		\bibitem [{\citenamefont {Basov}\ \emph {et~al.}(2017)\citenamefont {Basov},
			\citenamefont {Averitt},\ and\ \citenamefont {Hsieh}}]{Basov2017}%
		\BibitemOpen
		\bibfield  {author} {\bibinfo {author} {\bibfnamefont {D.~N.}\ \bibnamefont
				{Basov}}, \bibinfo {author} {\bibfnamefont {R.~D.}\ \bibnamefont {Averitt}},\
			and\ \bibinfo {author} {\bibfnamefont {D.}~\bibnamefont {Hsieh}},\ }\bibfield
		{title} {\bibinfo {title} {Towards properties on demand in quantum
				materials},\ }\href {https://doi.org/10.1038/nmat5017} {\bibfield  {journal}
			{\bibinfo  {journal} {Nature Materials}\ }\textbf {\bibinfo {volume} {16}},\
			\bibinfo {pages} {1077} (\bibinfo {year} {2017})}\BibitemShut {NoStop}%
		\bibitem [{\citenamefont {Giannetti}\ \emph {et~al.}(2016)\citenamefont
			{Giannetti}, \citenamefont {Capone}, \citenamefont {Fausti}, \citenamefont
			{Fabrizio}, \citenamefont {Parmigiani},\ and\ \citenamefont
			{Mihailovic}}]{Giannetti2016}%
		\BibitemOpen
		\bibfield  {author} {\bibinfo {author} {\bibfnamefont {C.}~\bibnamefont
				{Giannetti}}, \bibinfo {author} {\bibfnamefont {M.}~\bibnamefont {Capone}},
			\bibinfo {author} {\bibfnamefont {D.}~\bibnamefont {Fausti}}, \bibinfo
			{author} {\bibfnamefont {M.}~\bibnamefont {Fabrizio}}, \bibinfo {author}
			{\bibfnamefont {F.}~\bibnamefont {Parmigiani}},\ and\ \bibinfo {author}
			{\bibfnamefont {D.}~\bibnamefont {Mihailovic}},\ }\bibfield  {title}
		{\bibinfo {title} {Ultrafast optical spectroscopy of strongly correlated
				materials and high-temperature superconductors: a non-equilibrium approach},\
		}\href {https://doi.org/10.1080/00018732.2016.1194044} {\bibfield  {journal}
			{\bibinfo  {journal} {Advances in Physics}\ }\textbf {\bibinfo {volume}
				{65}},\ \bibinfo {pages} {58} (\bibinfo {year} {2016})}\BibitemShut {NoStop}%
		\bibitem [{\citenamefont {de~la Torre}\ \emph {et~al.}(2021)\citenamefont
			{de~la Torre}, \citenamefont {Kennes}, \citenamefont {Claassen},
			\citenamefont {Gerber}, \citenamefont {McIver},\ and\ \citenamefont
			{Sentef}}]{delaTorre2021}%
		\BibitemOpen
		\bibfield  {author} {\bibinfo {author} {\bibfnamefont {A.}~\bibnamefont
				{de~la Torre}}, \bibinfo {author} {\bibfnamefont {D.~M.}\ \bibnamefont
				{Kennes}}, \bibinfo {author} {\bibfnamefont {M.}~\bibnamefont {Claassen}},
			\bibinfo {author} {\bibfnamefont {S.}~\bibnamefont {Gerber}}, \bibinfo
			{author} {\bibfnamefont {J.~W.}\ \bibnamefont {McIver}},\ and\ \bibinfo
			{author} {\bibfnamefont {M.~A.}\ \bibnamefont {Sentef}},\ }\bibfield  {title}
		{\bibinfo {title} {{Colloquium: Nonthermal pathways to ultrafast control in
					quantum materials}},\ }\href {https://doi.org/10.1103/RevModPhys.93.041002}
		{\bibfield  {journal} {\bibinfo  {journal} {Rev. Mod. Phys.}\ }\textbf
			{\bibinfo {volume} {93}},\ \bibinfo {pages} {041002} (\bibinfo {year}
			{2021})}\BibitemShut {NoStop}%
		\bibitem [{\citenamefont {Frisk~Kockum}\ \emph {et~al.}(2019)\citenamefont
			{Frisk~Kockum}, \citenamefont {Miranowicz}, \citenamefont {De~Liberato},
			\citenamefont {Savasta},\ and\ \citenamefont {Nori}}]{FriskKockum2019}%
		\BibitemOpen
		\bibfield  {author} {\bibinfo {author} {\bibfnamefont {A.}~\bibnamefont
				{Frisk~Kockum}}, \bibinfo {author} {\bibfnamefont {A.}~\bibnamefont
				{Miranowicz}}, \bibinfo {author} {\bibfnamefont {S.}~\bibnamefont
				{De~Liberato}}, \bibinfo {author} {\bibfnamefont {S.}~\bibnamefont
				{Savasta}},\ and\ \bibinfo {author} {\bibfnamefont {F.}~\bibnamefont
				{Nori}},\ }\bibfield  {title} {\bibinfo {title} {Ultrastrong coupling between
				light and matter},\ }\href {https://doi.org/10.1038/s42254-018-0006-2}
		{\bibfield  {journal} {\bibinfo  {journal} {Nature Reviews Physics}\ }\textbf
			{\bibinfo {volume} {1}},\ \bibinfo {pages} {19} (\bibinfo {year}
			{2019})}\BibitemShut {NoStop}%
		\bibitem [{\citenamefont {Flick}\ \emph {et~al.}(2017)\citenamefont {Flick},
			\citenamefont {Ruggenthaler}, \citenamefont {Appel},\ and\ \citenamefont
			{Rubio}}]{Flick2017}%
		\BibitemOpen
		\bibfield  {author} {\bibinfo {author} {\bibfnamefont {J.}~\bibnamefont
				{Flick}}, \bibinfo {author} {\bibfnamefont {M.}~\bibnamefont {Ruggenthaler}},
			\bibinfo {author} {\bibfnamefont {H.}~\bibnamefont {Appel}},\ and\ \bibinfo
			{author} {\bibfnamefont {A.}~\bibnamefont {Rubio}},\ }\bibfield  {title}
		{\bibinfo {title} {Atoms and molecules in cavities, from weak to strong
				coupling in quantum-electrodynamics (qed) chemistry},\ }\href
		{https://doi.org/10.1073/pnas.1615509114} {\bibfield  {journal} {\bibinfo
				{journal} {Proceedings of the National Academy of Sciences}\ }\textbf
			{\bibinfo {volume} {114}},\ \bibinfo {pages} {3026} (\bibinfo {year}
			{2017})}\BibitemShut {NoStop}%
		\bibitem [{\citenamefont {Fregoni}\ \emph {et~al.}(2022)\citenamefont
			{Fregoni}, \citenamefont {Garcia-Vidal},\ and\ \citenamefont
			{Feist}}]{Fregoni2022}%
		\BibitemOpen
		\bibfield  {author} {\bibinfo {author} {\bibfnamefont {J.}~\bibnamefont
				{Fregoni}}, \bibinfo {author} {\bibfnamefont {F.~J.}\ \bibnamefont
				{Garcia-Vidal}},\ and\ \bibinfo {author} {\bibfnamefont {J.}~\bibnamefont
				{Feist}},\ }\bibfield  {title} {\bibinfo {title} {Theoretical challenges in
				polaritonic chemistry},\ }\href
		{https://doi.org/10.1021/acsphotonics.1c01749} {\bibfield  {journal}
			{\bibinfo  {journal} {ACS Photonics}\ }\textbf {\bibinfo {volume} {9}},\
			\bibinfo {pages} {1096} (\bibinfo {year} {2022})}\BibitemShut {NoStop}%
		\bibitem [{\citenamefont {Schäfer}\ \emph {et~al.}(2019)\citenamefont
			{Schäfer}, \citenamefont {Ruggenthaler}, \citenamefont {Appel},\ and\
			\citenamefont {Rubio}}]{Schaefer2019}%
		\BibitemOpen
		\bibfield  {author} {\bibinfo {author} {\bibfnamefont {C.}~\bibnamefont
				{Schäfer}}, \bibinfo {author} {\bibfnamefont {M.}~\bibnamefont
				{Ruggenthaler}}, \bibinfo {author} {\bibfnamefont {H.}~\bibnamefont
				{Appel}},\ and\ \bibinfo {author} {\bibfnamefont {A.}~\bibnamefont {Rubio}},\
		}\bibfield  {title} {\bibinfo {title} {Modification of excitation and charge
				transfer in cavity quantum-electrodynamical chemistry},\ }\href
		{https://doi.org/10.1073/pnas.1814178116} {\bibfield  {journal} {\bibinfo
				{journal} {Proceedings of the National Academy of Sciences}\ }\textbf
			{\bibinfo {volume} {116}},\ \bibinfo {pages} {4883} (\bibinfo {year}
			{2019})}\BibitemShut {NoStop}%
		\bibitem [{\citenamefont {Schlawin}\ \emph {et~al.}(2022)\citenamefont
			{Schlawin}, \citenamefont {Kennes},\ and\ \citenamefont
			{Sentef}}]{Schlawin2022}%
		\BibitemOpen
		\bibfield  {author} {\bibinfo {author} {\bibfnamefont {F.}~\bibnamefont
				{Schlawin}}, \bibinfo {author} {\bibfnamefont {D.~M.}\ \bibnamefont
				{Kennes}},\ and\ \bibinfo {author} {\bibfnamefont {M.~A.}\ \bibnamefont
				{Sentef}},\ }\bibfield  {title} {\bibinfo {title} {Cavity quantum
				materials},\ }\href {https://doi.org/10.1063/5.0083825} {\bibfield  {journal}
			{\bibinfo  {journal} {Applied Physics Reviews}\ }\textbf {\bibinfo {volume}
				{9}},\ \bibinfo {pages} {011312} (\bibinfo {year} {2022})}\BibitemShut
		{NoStop}%
		\bibitem [{\citenamefont {Scalari}\ \emph {et~al.}(2012)\citenamefont
			{Scalari}, \citenamefont {Maissen}, \citenamefont {Turčinková},
			\citenamefont {Hagenmüller}, \citenamefont {Liberato}, \citenamefont
			{Ciuti}, \citenamefont {Reichl}, \citenamefont {Schuh}, \citenamefont
			{Wegscheider}, \citenamefont {Beck},\ and\ \citenamefont
			{Faist}}]{Scalari2012}%
		\BibitemOpen
		\bibfield  {author} {\bibinfo {author} {\bibfnamefont {G.}~\bibnamefont
				{Scalari}}, \bibinfo {author} {\bibfnamefont {C.}~\bibnamefont {Maissen}},
			\bibinfo {author} {\bibfnamefont {D.}~\bibnamefont {Turčinková}}, \bibinfo
			{author} {\bibfnamefont {D.}~\bibnamefont {Hagenmüller}}, \bibinfo {author}
			{\bibfnamefont {S.~D.}\ \bibnamefont {Liberato}}, \bibinfo {author}
			{\bibfnamefont {C.}~\bibnamefont {Ciuti}}, \bibinfo {author} {\bibfnamefont
				{C.}~\bibnamefont {Reichl}}, \bibinfo {author} {\bibfnamefont
				{D.}~\bibnamefont {Schuh}}, \bibinfo {author} {\bibfnamefont
				{W.}~\bibnamefont {Wegscheider}}, \bibinfo {author} {\bibfnamefont
				{M.}~\bibnamefont {Beck}},\ and\ \bibinfo {author} {\bibfnamefont
				{J.}~\bibnamefont {Faist}},\ }\bibfield  {title} {\bibinfo {title}
			{Ultrastrong coupling of the cyclotron transition of a 2d electron gas to a
				{TH}z metamaterial},\ }\href {https://doi.org/10.1126/science.1216022}
		{\bibfield  {journal} {\bibinfo  {journal} {Science}\ }\textbf {\bibinfo
				{volume} {335}},\ \bibinfo {pages} {1323} (\bibinfo {year}
			{2012})}\BibitemShut {NoStop}%
		\bibitem [{\citenamefont {Zhang}\ \emph {et~al.}(2016)\citenamefont {Zhang},
			\citenamefont {Lou}, \citenamefont {Li}, \citenamefont {Reno}, \citenamefont
			{Pan}, \citenamefont {Watson}, \citenamefont {Manfra},\ and\ \citenamefont
			{Kono}}]{Zhang2016}%
		\BibitemOpen
		\bibfield  {author} {\bibinfo {author} {\bibfnamefont {Q.}~\bibnamefont
				{Zhang}}, \bibinfo {author} {\bibfnamefont {M.}~\bibnamefont {Lou}}, \bibinfo
			{author} {\bibfnamefont {X.}~\bibnamefont {Li}}, \bibinfo {author}
			{\bibfnamefont {J.~L.}\ \bibnamefont {Reno}}, \bibinfo {author}
			{\bibfnamefont {W.}~\bibnamefont {Pan}}, \bibinfo {author} {\bibfnamefont
				{J.~D.}\ \bibnamefont {Watson}}, \bibinfo {author} {\bibfnamefont {M.~J.}\
				\bibnamefont {Manfra}},\ and\ \bibinfo {author} {\bibfnamefont
				{J.}~\bibnamefont {Kono}},\ }\bibfield  {title} {\bibinfo {title} {Collective
				non-perturbative coupling of 2{D} electrons with high-quality-factor
				terahertz cavity photons},\ }\href {https://doi.org/10.1038/nphys3850}
		{\bibfield  {journal} {\bibinfo  {journal} {Nature Physics}\ }\textbf
			{\bibinfo {volume} {12}},\ \bibinfo {pages} {1005} (\bibinfo {year}
			{2016})}\BibitemShut {NoStop}%
		\bibitem [{\citenamefont {Thomas}\ \emph {et~al.}(2019)\citenamefont {Thomas},
			\citenamefont {Devaux}, \citenamefont {Nagarajan}, \citenamefont {Chervy},
			\citenamefont {Seidel}, \citenamefont {Hagenmüller}, \citenamefont
			{Schütz}, \citenamefont {Schachenmayer}, \citenamefont {Genet},
			\citenamefont {Pupillo},\ and\ \citenamefont {Ebbesen}}]{Thomas2019}%
		\BibitemOpen
		\bibfield  {author} {\bibinfo {author} {\bibfnamefont {A.}~\bibnamefont
				{Thomas}}, \bibinfo {author} {\bibfnamefont {E.}~\bibnamefont {Devaux}},
			\bibinfo {author} {\bibfnamefont {K.}~\bibnamefont {Nagarajan}}, \bibinfo
			{author} {\bibfnamefont {T.}~\bibnamefont {Chervy}}, \bibinfo {author}
			{\bibfnamefont {M.}~\bibnamefont {Seidel}}, \bibinfo {author} {\bibfnamefont
				{D.}~\bibnamefont {Hagenmüller}}, \bibinfo {author} {\bibfnamefont
				{S.}~\bibnamefont {Schütz}}, \bibinfo {author} {\bibfnamefont
				{J.}~\bibnamefont {Schachenmayer}}, \bibinfo {author} {\bibfnamefont
				{C.}~\bibnamefont {Genet}}, \bibinfo {author} {\bibfnamefont
				{G.}~\bibnamefont {Pupillo}},\ and\ \bibinfo {author} {\bibfnamefont {T.~W.}\
				\bibnamefont {Ebbesen}},\ }\href {https://doi.org/10.48550/ARXIV.1911.01459}
		{\bibinfo {title} {Exploring superconductivity under strong coupling with the
				vacuum electromagnetic field}} (\bibinfo {year} {2019})\BibitemShut {NoStop}%
		\bibitem [{\citenamefont {Thomas}\ \emph {et~al.}(2021)\citenamefont {Thomas},
			\citenamefont {Devaux}, \citenamefont {Nagarajan}, \citenamefont {Rogez},
			\citenamefont {Seidel}, \citenamefont {Richard}, \citenamefont {Genet},
			\citenamefont {Drillon},\ and\ \citenamefont {Ebbesen}}]{Thomas2021}%
		\BibitemOpen
		\bibfield  {author} {\bibinfo {author} {\bibfnamefont {A.}~\bibnamefont
				{Thomas}}, \bibinfo {author} {\bibfnamefont {E.}~\bibnamefont {Devaux}},
			\bibinfo {author} {\bibfnamefont {K.}~\bibnamefont {Nagarajan}}, \bibinfo
			{author} {\bibfnamefont {G.}~\bibnamefont {Rogez}}, \bibinfo {author}
			{\bibfnamefont {M.}~\bibnamefont {Seidel}}, \bibinfo {author} {\bibfnamefont
				{F.}~\bibnamefont {Richard}}, \bibinfo {author} {\bibfnamefont
				{C.}~\bibnamefont {Genet}}, \bibinfo {author} {\bibfnamefont
				{M.}~\bibnamefont {Drillon}},\ and\ \bibinfo {author} {\bibfnamefont {T.~W.}\
				\bibnamefont {Ebbesen}},\ }\bibfield  {title} {\bibinfo {title} {Large
				enhancement of ferromagnetism under a collective strong coupling of ybco
				nanoparticles},\ }\bibfield  {booktitle} {\emph {\bibinfo {booktitle} {Nano
					Letters}},\ }\href {https://doi.org/10.1021/acs.nanolett.1c00973} {\bibfield
			{journal} {\bibinfo  {journal} {Nano Letters}\ }\textbf {\bibinfo {volume}
				{21}},\ \bibinfo {pages} {4365} (\bibinfo {year} {2021})}\BibitemShut
		{NoStop}%
		\bibitem [{\citenamefont {Appugliese}\ \emph {et~al.}(2022)\citenamefont
			{Appugliese}, \citenamefont {Enkner}, \citenamefont {Paravicini-Bagliani},
			\citenamefont {Beck}, \citenamefont {Reichl}, \citenamefont {Wegscheider},
			\citenamefont {Scalari}, \citenamefont {Ciuti},\ and\ \citenamefont
			{Faist}}]{Appugliese2022}%
		\BibitemOpen
		\bibfield  {author} {\bibinfo {author} {\bibfnamefont {F.}~\bibnamefont
				{Appugliese}}, \bibinfo {author} {\bibfnamefont {J.}~\bibnamefont {Enkner}},
			\bibinfo {author} {\bibfnamefont {G.~L.}\ \bibnamefont
				{Paravicini-Bagliani}}, \bibinfo {author} {\bibfnamefont {M.}~\bibnamefont
				{Beck}}, \bibinfo {author} {\bibfnamefont {C.}~\bibnamefont {Reichl}},
			\bibinfo {author} {\bibfnamefont {W.}~\bibnamefont {Wegscheider}}, \bibinfo
			{author} {\bibfnamefont {G.}~\bibnamefont {Scalari}}, \bibinfo {author}
			{\bibfnamefont {C.}~\bibnamefont {Ciuti}},\ and\ \bibinfo {author}
			{\bibfnamefont {J.}~\bibnamefont {Faist}},\ }\bibfield  {title} {\bibinfo
			{title} {Breakdown of topological protection by cavity vacuum fields in the
				integer quantum hall effect},\ }\href
		{https://doi.org/10.1126/science.abl5818} {\bibfield  {journal} {\bibinfo
				{journal} {Science}\ }\textbf {\bibinfo {volume} {375}},\ \bibinfo {pages}
			{1030} (\bibinfo {year} {2022})}\BibitemShut {NoStop}%
		\bibitem [{\citenamefont {Sentef}\ \emph {et~al.}(2018)\citenamefont {Sentef},
			\citenamefont {Ruggenthaler},\ and\ \citenamefont {Rubio}}]{Sentef2018}%
		\BibitemOpen
		\bibfield  {author} {\bibinfo {author} {\bibfnamefont {M.~A.}\ \bibnamefont
				{Sentef}}, \bibinfo {author} {\bibfnamefont {M.}~\bibnamefont
				{Ruggenthaler}},\ and\ \bibinfo {author} {\bibfnamefont {A.}~\bibnamefont
				{Rubio}},\ }\bibfield  {title} {\bibinfo {title} {Cavity
				quantum-electrodynamical polaritonically enhanced electron-phonon coupling
				and its influence on superconductivity},\ }\href
		{https://doi.org/10.1126/sciadv.aau6969} {\bibfield  {journal} {\bibinfo
				{journal} {Science Advances}\ }\textbf {\bibinfo {volume} {4}},\ \bibinfo
			{pages} {eaau6969} (\bibinfo {year} {2018})}\BibitemShut {NoStop}%
		\bibitem [{\citenamefont {Schlawin}\ \emph {et~al.}(2019)\citenamefont
			{Schlawin}, \citenamefont {Cavalleri},\ and\ \citenamefont
			{Jaksch}}]{Schlawin2019}%
		\BibitemOpen
		\bibfield  {author} {\bibinfo {author} {\bibfnamefont {F.}~\bibnamefont
				{Schlawin}}, \bibinfo {author} {\bibfnamefont {A.}~\bibnamefont
				{Cavalleri}},\ and\ \bibinfo {author} {\bibfnamefont {D.}~\bibnamefont
				{Jaksch}},\ }\bibfield  {title} {\bibinfo {title} {Cavity-mediated
				electron-photon superconductivity},\ }\href
		{https://doi.org/10.1103/PhysRevLett.122.133602} {\bibfield  {journal}
			{\bibinfo  {journal} {Phys. Rev. Lett.}\ }\textbf {\bibinfo {volume} {122}},\
			\bibinfo {pages} {133602} (\bibinfo {year} {2019})}\BibitemShut {NoStop}%
		\bibitem [{\citenamefont {Li}\ and\ \citenamefont {Eckstein}(2020)}]{Li2020b}%
		\BibitemOpen
		\bibfield  {author} {\bibinfo {author} {\bibfnamefont {J.}~\bibnamefont
				{Li}}\ and\ \bibinfo {author} {\bibfnamefont {M.}~\bibnamefont {Eckstein}},\
		}\bibfield  {title} {\bibinfo {title} {Manipulating intertwined orders in
				solids with quantum light},\ }\href
		{https://doi.org/10.1103/PhysRevLett.125.217402} {\bibfield  {journal}
			{\bibinfo  {journal} {Phys. Rev. Lett.}\ }\textbf {\bibinfo {volume} {125}},\
			\bibinfo {pages} {217402} (\bibinfo {year} {2020})}\BibitemShut {NoStop}%
		\bibitem [{\citenamefont {Wang}\ \emph {et~al.}(2019)\citenamefont {Wang},
			\citenamefont {Ronca},\ and\ \citenamefont {Sentef}}]{Wang2019}%
		\BibitemOpen
		\bibfield  {author} {\bibinfo {author} {\bibfnamefont {X.}~\bibnamefont
				{Wang}}, \bibinfo {author} {\bibfnamefont {E.}~\bibnamefont {Ronca}},\ and\
			\bibinfo {author} {\bibfnamefont {M.~A.}\ \bibnamefont {Sentef}},\ }\bibfield
		{title} {\bibinfo {title} {Cavity quantum electrodynamical chern insulator:
				Towards light-induced quantized anomalous hall effect in graphene},\ }\href
		{https://doi.org/10.1103/PhysRevB.99.235156} {\bibfield  {journal} {\bibinfo
				{journal} {Phys. Rev. B}\ }\textbf {\bibinfo {volume} {99}},\ \bibinfo
			{pages} {235156} (\bibinfo {year} {2019})}\BibitemShut {NoStop}%
		\bibitem [{\citenamefont {Dmytruk}\ and\ \citenamefont
			{Schirò}(2022)}]{Dmytruk2022}%
		\BibitemOpen
		\bibfield  {author} {\bibinfo {author} {\bibfnamefont {O.}~\bibnamefont
				{Dmytruk}}\ and\ \bibinfo {author} {\bibfnamefont {M.}~\bibnamefont
				{Schirò}},\ }\href@noop {} {\bibinfo {title} {Controlling topological phases
				of matter with quantum light}} (\bibinfo {year} {2022}),\ \Eprint
		{https://arxiv.org/abs/2204.05922} {arXiv:2204.05922 [cond-mat.mes-hall]}
		\BibitemShut {NoStop}%
		\bibitem [{\citenamefont {Kiffner}\ \emph {et~al.}(2019)\citenamefont
			{Kiffner}, \citenamefont {Coulthard}, \citenamefont {Schlawin}, \citenamefont
			{Ardavan},\ and\ \citenamefont {Jaksch}}]{Kiffner2019}%
		\BibitemOpen
		\bibfield  {author} {\bibinfo {author} {\bibfnamefont {M.}~\bibnamefont
				{Kiffner}}, \bibinfo {author} {\bibfnamefont {J.~R.}\ \bibnamefont
				{Coulthard}}, \bibinfo {author} {\bibfnamefont {F.}~\bibnamefont {Schlawin}},
			\bibinfo {author} {\bibfnamefont {A.}~\bibnamefont {Ardavan}},\ and\ \bibinfo
			{author} {\bibfnamefont {D.}~\bibnamefont {Jaksch}},\ }\bibfield  {title}
		{\bibinfo {title} {Manipulating quantum materials with quantum light},\
		}\href {https://doi.org/10.1103/PhysRevB.99.085116} {\bibfield  {journal}
			{\bibinfo  {journal} {Phys. Rev. B}\ }\textbf {\bibinfo {volume} {99}},\
			\bibinfo {pages} {085116} (\bibinfo {year} {2019})}\BibitemShut {NoStop}%
		\bibitem [{\citenamefont {Sentef}\ \emph {et~al.}(2020)\citenamefont {Sentef},
			\citenamefont {Li}, \citenamefont {K\"unzel},\ and\ \citenamefont
			{Eckstein}}]{Sentef2020}%
		\BibitemOpen
		\bibfield  {author} {\bibinfo {author} {\bibfnamefont {M.~A.}\ \bibnamefont
				{Sentef}}, \bibinfo {author} {\bibfnamefont {J.}~\bibnamefont {Li}}, \bibinfo
			{author} {\bibfnamefont {F.}~\bibnamefont {K\"unzel}},\ and\ \bibinfo
			{author} {\bibfnamefont {M.}~\bibnamefont {Eckstein}},\ }\bibfield  {title}
		{\bibinfo {title} {Quantum to classical crossover of floquet engineering in
				correlated quantum systems},\ }\href
		{https://doi.org/10.1103/PhysRevResearch.2.033033} {\bibfield  {journal}
			{\bibinfo  {journal} {Phys. Rev. Research}\ }\textbf {\bibinfo {volume}
				{2}},\ \bibinfo {pages} {033033} (\bibinfo {year} {2020})}\BibitemShut
		{NoStop}%
		\bibitem [{\citenamefont {M\"uller}\ \emph {et~al.}(2022)\citenamefont
			{M\"uller}, \citenamefont {Eckstein},\ and\ \citenamefont
			{Kusminskiy}}]{MuellerH2022}%
		\BibitemOpen
		\bibfield  {author} {\bibinfo {author} {\bibfnamefont {H.}~\bibnamefont
				{M\"uller}}, \bibinfo {author} {\bibfnamefont {M.}~\bibnamefont {Eckstein}},\
			and\ \bibinfo {author} {\bibfnamefont {S.~V.}\ \bibnamefont {Kusminskiy}},\
		}\href@noop {} {\bibinfo {title} {Control of yu-shiba-rusinov states through
				a bosonic mode}} (\bibinfo {year} {2022}),\ \Eprint
		{https://arxiv.org/abs/2207.14180} {arXiv:2207.14180 [cond-mat.str-el]}
		\BibitemShut {NoStop}%
		\bibitem [{\citenamefont {Curtis}\ \emph {et~al.}(2022)\citenamefont {Curtis},
			\citenamefont {Grankin}, \citenamefont {Poniatowski}, \citenamefont
			{Galitski}, \citenamefont {Narang},\ and\ \citenamefont
			{Demler}}]{Curtis2022}%
		\BibitemOpen
		\bibfield  {author} {\bibinfo {author} {\bibfnamefont {J.~B.}\ \bibnamefont
				{Curtis}}, \bibinfo {author} {\bibfnamefont {A.}~\bibnamefont {Grankin}},
			\bibinfo {author} {\bibfnamefont {N.~R.}\ \bibnamefont {Poniatowski}},
			\bibinfo {author} {\bibfnamefont {V.~M.}\ \bibnamefont {Galitski}}, \bibinfo
			{author} {\bibfnamefont {P.}~\bibnamefont {Narang}},\ and\ \bibinfo {author}
			{\bibfnamefont {E.}~\bibnamefont {Demler}},\ }\bibfield  {title} {\bibinfo
			{title} {Cavity magnon-polaritons in cuprate parent compounds},\ }\href
		{https://doi.org/10.1103/PhysRevResearch.4.013101} {\bibfield  {journal}
			{\bibinfo  {journal} {Phys. Rev. Research}\ }\textbf {\bibinfo {volume}
				{4}},\ \bibinfo {pages} {013101} (\bibinfo {year} {2022})}\BibitemShut
		{NoStop}%
		\bibitem [{\citenamefont {Chiocchetta}\ \emph {et~al.}(2021)\citenamefont
			{Chiocchetta}, \citenamefont {Kiese}, \citenamefont {Zelle}, \citenamefont
			{Piazza},\ and\ \citenamefont {Diehl}}]{Chiocchetta2021}%
		\BibitemOpen
		\bibfield  {author} {\bibinfo {author} {\bibfnamefont {A.}~\bibnamefont
				{Chiocchetta}}, \bibinfo {author} {\bibfnamefont {D.}~\bibnamefont {Kiese}},
			\bibinfo {author} {\bibfnamefont {C.~P.}\ \bibnamefont {Zelle}}, \bibinfo
			{author} {\bibfnamefont {F.}~\bibnamefont {Piazza}},\ and\ \bibinfo {author}
			{\bibfnamefont {S.}~\bibnamefont {Diehl}},\ }\bibfield  {title} {\bibinfo
			{title} {Cavity-induced quantum spin liquids},\ }\href
		{https://doi.org/10.1038/s41467-021-26076-3} {\bibfield  {journal} {\bibinfo
				{journal} {Nature Communications}\ }\textbf {\bibinfo {volume} {12}},\
			\bibinfo {pages} {5901} (\bibinfo {year} {2021})}\BibitemShut {NoStop}%
		\bibitem [{\citenamefont {Mazza}\ and\ \citenamefont
			{Georges}(2019)}]{Mazza2019}%
		\BibitemOpen
		\bibfield  {author} {\bibinfo {author} {\bibfnamefont {G.}~\bibnamefont
				{Mazza}}\ and\ \bibinfo {author} {\bibfnamefont {A.}~\bibnamefont
				{Georges}},\ }\bibfield  {title} {\bibinfo {title} {Superradiant quantum
				materials},\ }\href {https://doi.org/10.1103/PhysRevLett.122.017401}
		{\bibfield  {journal} {\bibinfo  {journal} {Phys. Rev. Lett.}\ }\textbf
			{\bibinfo {volume} {122}},\ \bibinfo {pages} {017401} (\bibinfo {year}
			{2019})}\BibitemShut {NoStop}%
		\bibitem [{\citenamefont {Lenk}\ and\ \citenamefont
			{Eckstein}(2020)}]{Lenk2020}%
		\BibitemOpen
		\bibfield  {author} {\bibinfo {author} {\bibfnamefont {K.}~\bibnamefont
				{Lenk}}\ and\ \bibinfo {author} {\bibfnamefont {M.}~\bibnamefont
				{Eckstein}},\ }\bibfield  {title} {\bibinfo {title} {Collective excitations
				of the {U}(1)-symmetric exciton insulator in a cavity},\ }\bibfield
		{journal} {\bibinfo  {journal} {Physical Review B}\ }\textbf {\bibinfo
			{volume} {102}},\ \href {https://doi.org/10.1103/physrevb.102.205129}
		{10.1103/physrevb.102.205129} (\bibinfo {year} {2020})\BibitemShut {NoStop}%
		\bibitem [{\citenamefont {Ashida}\ \emph {et~al.}(2020)\citenamefont {Ashida},
			\citenamefont {\ifmmode \dot{I}\else \.{I}\fi{}mamo\ifmmode~\breve{g}\else
				\u{g}\fi{}lu}, \citenamefont {Faist}, \citenamefont {Jaksch}, \citenamefont
			{Cavalleri},\ and\ \citenamefont {Demler}}]{Ashida2020}%
		\BibitemOpen
		\bibfield  {author} {\bibinfo {author} {\bibfnamefont {Y.}~\bibnamefont
				{Ashida}}, \bibinfo {author} {\bibfnamefont {A.~m.~c.}\ \bibnamefont
				{\ifmmode \dot{I}\else \.{I}\fi{}mamo\ifmmode~\breve{g}\else \u{g}\fi{}lu}},
			\bibinfo {author} {\bibfnamefont {J.}~\bibnamefont {Faist}}, \bibinfo
			{author} {\bibfnamefont {D.}~\bibnamefont {Jaksch}}, \bibinfo {author}
			{\bibfnamefont {A.}~\bibnamefont {Cavalleri}},\ and\ \bibinfo {author}
			{\bibfnamefont {E.}~\bibnamefont {Demler}},\ }\bibfield  {title} {\bibinfo
			{title} {Quantum electrodynamic control of matter: Cavity-enhanced
				ferroelectric phase transition},\ }\href
		{https://doi.org/10.1103/PhysRevX.10.041027} {\bibfield  {journal} {\bibinfo
				{journal} {Phys. Rev. X}\ }\textbf {\bibinfo {volume} {10}},\ \bibinfo
			{pages} {041027} (\bibinfo {year} {2020})}\BibitemShut {NoStop}%
		\bibitem [{\citenamefont {Latini}\ \emph {et~al.}(2021)\citenamefont {Latini},
			\citenamefont {Shin}, \citenamefont {Sato}, \citenamefont {Schäfer},
			\citenamefont {Giovannini}, \citenamefont {Hübener},\ and\ \citenamefont
			{Rubio}}]{Latini2021}%
		\BibitemOpen
		\bibfield  {author} {\bibinfo {author} {\bibfnamefont {S.}~\bibnamefont
				{Latini}}, \bibinfo {author} {\bibfnamefont {D.}~\bibnamefont {Shin}},
			\bibinfo {author} {\bibfnamefont {S.~A.}\ \bibnamefont {Sato}}, \bibinfo
			{author} {\bibfnamefont {C.}~\bibnamefont {Schäfer}}, \bibinfo {author}
			{\bibfnamefont {U.~D.}\ \bibnamefont {Giovannini}}, \bibinfo {author}
			{\bibfnamefont {H.}~\bibnamefont {Hübener}},\ and\ \bibinfo {author}
			{\bibfnamefont {A.}~\bibnamefont {Rubio}},\ }\bibfield  {title} {\bibinfo
			{title} {The ferroelectric photo ground state of srtio$_3$: Cavity materials
				engineering},\ }\href {https://doi.org/10.1073/pnas.2105618118} {\bibfield
			{journal} {\bibinfo  {journal} {Proceedings of the National Academy of
					Sciences}\ }\textbf {\bibinfo {volume} {118}},\ \bibinfo {pages}
			{e2105618118} (\bibinfo {year} {2021})}\BibitemShut {NoStop}%
		\bibitem [{\citenamefont {Schuler}\ \emph {et~al.}(2020)\citenamefont
			{Schuler}, \citenamefont {Bernardis}, \citenamefont {Läuchli},\ and\
			\citenamefont {Rabl}}]{Schuler2020}%
		\BibitemOpen
		\bibfield  {author} {\bibinfo {author} {\bibfnamefont {M.}~\bibnamefont
				{Schuler}}, \bibinfo {author} {\bibfnamefont {D.~D.}\ \bibnamefont
				{Bernardis}}, \bibinfo {author} {\bibfnamefont {A.~M.}\ \bibnamefont
				{Läuchli}},\ and\ \bibinfo {author} {\bibfnamefont {P.}~\bibnamefont
				{Rabl}},\ }\bibfield  {title} {\bibinfo {title} {{The vacua of dipolar cavity
					quantum electrodynamics}},\ }\href
		{https://doi.org/10.21468/SciPostPhys.9.5.066} {\bibfield  {journal}
			{\bibinfo  {journal} {SciPost Phys.}\ }\textbf {\bibinfo {volume} {9}},\
			\bibinfo {pages} {066} (\bibinfo {year} {2020})}\BibitemShut {NoStop}%
		\bibitem [{\citenamefont {De~Bernardis}\ \emph {et~al.}(2018)\citenamefont
			{De~Bernardis}, \citenamefont {Jaako},\ and\ \citenamefont
			{Rabl}}]{Bernardis2018}%
		\BibitemOpen
		\bibfield  {author} {\bibinfo {author} {\bibfnamefont {D.}~\bibnamefont
				{De~Bernardis}}, \bibinfo {author} {\bibfnamefont {T.}~\bibnamefont
				{Jaako}},\ and\ \bibinfo {author} {\bibfnamefont {P.}~\bibnamefont {Rabl}},\
		}\bibfield  {title} {\bibinfo {title} {Cavity quantum electrodynamics in the
				nonperturbative regime},\ }\href {https://doi.org/10.1103/PhysRevA.97.043820}
		{\bibfield  {journal} {\bibinfo  {journal} {Phys. Rev. A}\ }\textbf {\bibinfo
				{volume} {97}},\ \bibinfo {pages} {043820} (\bibinfo {year}
			{2018})}\BibitemShut {NoStop}%
		\bibitem [{\citenamefont {Keeling}(2007)}]{Keeling2007}%
		\BibitemOpen
		\bibfield  {author} {\bibinfo {author} {\bibfnamefont {J.}~\bibnamefont
				{Keeling}},\ }\bibfield  {title} {\bibinfo {title} {Coulomb interactions,
				gauge invariance, and phase transitions of the dicke model},\ }\href
		{https://doi.org/10.1088/0953-8984/19/29/295213} {\bibfield  {journal}
			{\bibinfo  {journal} {Journal of Physics: Condensed Matter}\ }\textbf
			{\bibinfo {volume} {19}},\ \bibinfo {pages} {295213} (\bibinfo {year}
			{2007})}\BibitemShut {NoStop}%
		\bibitem [{\citenamefont {Lenk}\ \emph {et~al.}(2022)\citenamefont {Lenk},
			\citenamefont {Li}, \citenamefont {Werner},\ and\ \citenamefont
			{Eckstein}}]{Lenk2022}%
		\BibitemOpen
		\bibfield  {author} {\bibinfo {author} {\bibfnamefont {K.}~\bibnamefont
				{Lenk}}, \bibinfo {author} {\bibfnamefont {J.}~\bibnamefont {Li}}, \bibinfo
			{author} {\bibfnamefont {P.}~\bibnamefont {Werner}},\ and\ \bibinfo {author}
			{\bibfnamefont {M.}~\bibnamefont {Eckstein}},\ }\href
		{https://doi.org/10.48550/ARXIV.2205.05559} {\bibinfo {title} {Collective
				theory for an interacting solid in a single-mode cavity}} (\bibinfo {year}
		{2022})\BibitemShut {NoStop}%
		\bibitem [{\citenamefont {Pilar}\ \emph {et~al.}(2020)\citenamefont {Pilar},
			\citenamefont {De~Bernardis},\ and\ \citenamefont {Rabl}}]{Pilar2020}%
		\BibitemOpen
		\bibfield  {author} {\bibinfo {author} {\bibfnamefont {P.}~\bibnamefont
				{Pilar}}, \bibinfo {author} {\bibfnamefont {D.}~\bibnamefont
				{De~Bernardis}},\ and\ \bibinfo {author} {\bibfnamefont {P.}~\bibnamefont
				{Rabl}},\ }\bibfield  {title} {\bibinfo {title} {Thermodynamics of
				ultrastrongly coupled light-matter systems},\ }\href
		{https://doi.org/10.22331/q-2020-09-28-335} {\bibfield  {journal} {\bibinfo
				{journal} {Quantum}\ }\textbf {\bibinfo {volume} {4}},\ \bibinfo {pages}
			{335} (\bibinfo {year} {2020})}\BibitemShut {NoStop}%
		\bibitem [{\citenamefont {Caldwell}\ \emph {et~al.}(2014)\citenamefont
			{Caldwell}, \citenamefont {Kretinin}, \citenamefont {Chen}, \citenamefont
			{Giannini}, \citenamefont {Fogler}, \citenamefont {Francescato},
			\citenamefont {Ellis}, \citenamefont {Tischler}, \citenamefont {Woods},
			\citenamefont {Giles}, \citenamefont {Hong}, \citenamefont {Watanabe},
			\citenamefont {Taniguchi}, \citenamefont {Maier},\ and\ \citenamefont
			{Novoselov}}]{Caldwell2014}%
		\BibitemOpen
		\bibfield  {author} {\bibinfo {author} {\bibfnamefont {J.~D.}\ \bibnamefont
				{Caldwell}}, \bibinfo {author} {\bibfnamefont {A.~V.}\ \bibnamefont
				{Kretinin}}, \bibinfo {author} {\bibfnamefont {Y.}~\bibnamefont {Chen}},
			\bibinfo {author} {\bibfnamefont {V.}~\bibnamefont {Giannini}}, \bibinfo
			{author} {\bibfnamefont {M.~M.}\ \bibnamefont {Fogler}}, \bibinfo {author}
			{\bibfnamefont {Y.}~\bibnamefont {Francescato}}, \bibinfo {author}
			{\bibfnamefont {C.~T.}\ \bibnamefont {Ellis}}, \bibinfo {author}
			{\bibfnamefont {J.~G.}\ \bibnamefont {Tischler}}, \bibinfo {author}
			{\bibfnamefont {C.~R.}\ \bibnamefont {Woods}}, \bibinfo {author}
			{\bibfnamefont {A.~J.}\ \bibnamefont {Giles}}, \bibinfo {author}
			{\bibfnamefont {M.}~\bibnamefont {Hong}}, \bibinfo {author} {\bibfnamefont
				{K.}~\bibnamefont {Watanabe}}, \bibinfo {author} {\bibfnamefont
				{T.}~\bibnamefont {Taniguchi}}, \bibinfo {author} {\bibfnamefont {S.~A.}\
				\bibnamefont {Maier}},\ and\ \bibinfo {author} {\bibfnamefont {K.~S.}\
				\bibnamefont {Novoselov}},\ }\bibfield  {title} {\bibinfo {title}
			{Sub-diffractional volume-confined polaritons in the natural hyperbolic
				material hexagonal boron nitride},\ }\href
		{https://doi.org/10.1038/ncomms6221} {\bibfield  {journal} {\bibinfo
				{journal} {Nature Communications}\ }\textbf {\bibinfo {volume} {5}},\
			\bibinfo {pages} {5221} (\bibinfo {year} {2014})}\BibitemShut {NoStop}%
		\bibitem [{\citenamefont {M\"uller}\ and\ \citenamefont
			{Burkard}(1979)}]{Mueller1979}%
		\BibitemOpen
		\bibfield  {author} {\bibinfo {author} {\bibfnamefont {K.~A.}\ \bibnamefont
				{M\"uller}}\ and\ \bibinfo {author} {\bibfnamefont {H.}~\bibnamefont
				{Burkard}},\ }\bibfield  {title} {\bibinfo {title} {Srtio$_{3}$: An intrinsic
				quantum paraelectric below 4 k},\ }\href
		{https://doi.org/10.1103/PhysRevB.19.3593} {\bibfield  {journal} {\bibinfo
				{journal} {Phys. Rev. B}\ }\textbf {\bibinfo {volume} {19}},\ \bibinfo
			{pages} {3593} (\bibinfo {year} {1979})}\BibitemShut {NoStop}%
		\bibitem [{\citenamefont {Andolina}\ \emph {et~al.}(2019)\citenamefont
			{Andolina}, \citenamefont {Pellegrino}, \citenamefont {Giovannetti},
			\citenamefont {MacDonald},\ and\ \citenamefont {Polini}}]{Andolina2019}%
		\BibitemOpen
		\bibfield  {author} {\bibinfo {author} {\bibfnamefont {G.~M.}\ \bibnamefont
				{Andolina}}, \bibinfo {author} {\bibfnamefont {F.~M.~D.}\ \bibnamefont
				{Pellegrino}}, \bibinfo {author} {\bibfnamefont {V.}~\bibnamefont
				{Giovannetti}}, \bibinfo {author} {\bibfnamefont {A.~H.}\ \bibnamefont
				{MacDonald}},\ and\ \bibinfo {author} {\bibfnamefont {M.}~\bibnamefont
				{Polini}},\ }\bibfield  {title} {\bibinfo {title} {Cavity quantum
				electrodynamics of strongly correlated electron systems: A no-go theorem for
				photon condensation},\ }\href {https://doi.org/10.1103/PhysRevB.100.121109}
		{\bibfield  {journal} {\bibinfo  {journal} {Phys. Rev. B}\ }\textbf {\bibinfo
				{volume} {100}},\ \bibinfo {pages} {121109} (\bibinfo {year}
			{2019})}\BibitemShut {NoStop}%
		\bibitem [{\citenamefont {Andolina}\ \emph {et~al.}(2020)\citenamefont
			{Andolina}, \citenamefont {Pellegrino}, \citenamefont {Giovannetti},
			\citenamefont {MacDonald},\ and\ \citenamefont {Polini}}]{Andolina2020}%
		\BibitemOpen
		\bibfield  {author} {\bibinfo {author} {\bibfnamefont {G.~M.}\ \bibnamefont
				{Andolina}}, \bibinfo {author} {\bibfnamefont {F.~M.~D.}\ \bibnamefont
				{Pellegrino}}, \bibinfo {author} {\bibfnamefont {V.}~\bibnamefont
				{Giovannetti}}, \bibinfo {author} {\bibfnamefont {A.~H.}\ \bibnamefont
				{MacDonald}},\ and\ \bibinfo {author} {\bibfnamefont {M.}~\bibnamefont
				{Polini}},\ }\bibfield  {title} {\bibinfo {title} {Theory of photon
				condensation in a spatially varying electromagnetic field},\ }\href
		{https://doi.org/10.1103/PhysRevB.102.125137} {\bibfield  {journal} {\bibinfo
				{journal} {Phys. Rev. B}\ }\textbf {\bibinfo {volume} {102}},\ \bibinfo
			{pages} {125137} (\bibinfo {year} {2020})}\BibitemShut {NoStop}%
		\bibitem [{\citenamefont {Metzner}\ and\ \citenamefont
			{Vollhardt}(1989)}]{Metzner1989}%
		\BibitemOpen
		\bibfield  {author} {\bibinfo {author} {\bibfnamefont {W.}~\bibnamefont
				{Metzner}}\ and\ \bibinfo {author} {\bibfnamefont {D.}~\bibnamefont
				{Vollhardt}},\ }\bibfield  {title} {\bibinfo {title} {Correlated lattice
				fermions in $d=\ensuremath{\infty}$ dimensions},\ }\href
		{https://doi.org/10.1103/PhysRevLett.62.324} {\bibfield  {journal} {\bibinfo
				{journal} {Phys. Rev. Lett.}\ }\textbf {\bibinfo {volume} {62}},\ \bibinfo
			{pages} {324} (\bibinfo {year} {1989})}\BibitemShut {NoStop}%
		\bibitem [{\citenamefont {Georges}\ \emph
			{et~al.}(1996{\natexlab{a}})\citenamefont {Georges}, \citenamefont {Kotliar},
			\citenamefont {Krauth},\ and\ \citenamefont {Rozenberg}}]{Georges1996}%
		\BibitemOpen
		\bibfield  {author} {\bibinfo {author} {\bibfnamefont {A.}~\bibnamefont
				{Georges}}, \bibinfo {author} {\bibfnamefont {G.}~\bibnamefont {Kotliar}},
			\bibinfo {author} {\bibfnamefont {W.}~\bibnamefont {Krauth}},\ and\ \bibinfo
			{author} {\bibfnamefont {M.~J.}\ \bibnamefont {Rozenberg}},\ }\bibfield
		{title} {\bibinfo {title} {Dynamical mean-field theory of strongly correlated
				fermion systems and the limit of infinite dimensions},\ }\href
		{https://doi.org/10.1103/RevModPhys.68.13} {\bibfield  {journal} {\bibinfo
				{journal} {Rev. Mod. Phys.}\ }\textbf {\bibinfo {volume} {68}},\ \bibinfo
			{pages} {13} (\bibinfo {year} {1996}{\natexlab{a}})}\BibitemShut {NoStop}%
		\bibitem [{\citenamefont {Sun}\ and\ \citenamefont {Kotliar}(2002)}]{Sun2002}%
		\BibitemOpen
		\bibfield  {author} {\bibinfo {author} {\bibfnamefont {P.}~\bibnamefont
				{Sun}}\ and\ \bibinfo {author} {\bibfnamefont {G.}~\bibnamefont {Kotliar}},\
		}\bibfield  {title} {\bibinfo {title} {Extended dynamical mean-field theory
				and $\mathrm{GW}$ method},\ }\href
		{https://doi.org/10.1103/PhysRevB.66.085120} {\bibfield  {journal} {\bibinfo
				{journal} {Phys. Rev. B}\ }\textbf {\bibinfo {volume} {66}},\ \bibinfo
			{pages} {085120} (\bibinfo {year} {2002})}\BibitemShut {NoStop}%
		\bibitem [{\citenamefont {Byczuk}\ and\ \citenamefont
			{Vollhardt}(2008)}]{Byczuk2008}%
		\BibitemOpen
		\bibfield  {author} {\bibinfo {author} {\bibfnamefont {K.}~\bibnamefont
				{Byczuk}}\ and\ \bibinfo {author} {\bibfnamefont {D.}~\bibnamefont
				{Vollhardt}},\ }\bibfield  {title} {\bibinfo {title} {Correlated bosons on a
				lattice: Dynamical mean-field theory for bose-einstein condensed and normal
				phases},\ }\href {https://doi.org/10.1103/PhysRevB.77.235106} {\bibfield
			{journal} {\bibinfo  {journal} {Phys. Rev. B}\ }\textbf {\bibinfo {volume}
				{77}},\ \bibinfo {pages} {235106} (\bibinfo {year} {2008})}\BibitemShut
		{NoStop}%
		\bibitem [{\citenamefont {Hu}\ and\ \citenamefont {Tong}(2009)}]{Hu2009}%
		\BibitemOpen
		\bibfield  {author} {\bibinfo {author} {\bibfnamefont {W.-J.}\ \bibnamefont
				{Hu}}\ and\ \bibinfo {author} {\bibfnamefont {N.-H.}\ \bibnamefont {Tong}},\
		}\bibfield  {title} {\bibinfo {title} {Dynamical mean-field theory for the
				bose-hubbard model},\ }\href {https://doi.org/10.1103/PhysRevB.80.245110}
		{\bibfield  {journal} {\bibinfo  {journal} {Phys. Rev. B}\ }\textbf {\bibinfo
				{volume} {80}},\ \bibinfo {pages} {245110} (\bibinfo {year}
			{2009})}\BibitemShut {NoStop}%
		\bibitem [{\citenamefont {Anders}\ \emph {et~al.}(2010)\citenamefont {Anders},
			\citenamefont {Gull}, \citenamefont {Pollet}, \citenamefont {Troyer},\ and\
			\citenamefont {Werner}}]{Anders2010}%
		\BibitemOpen
		\bibfield  {author} {\bibinfo {author} {\bibfnamefont {P.}~\bibnamefont
				{Anders}}, \bibinfo {author} {\bibfnamefont {E.}~\bibnamefont {Gull}},
			\bibinfo {author} {\bibfnamefont {L.}~\bibnamefont {Pollet}}, \bibinfo
			{author} {\bibfnamefont {M.}~\bibnamefont {Troyer}},\ and\ \bibinfo {author}
			{\bibfnamefont {P.}~\bibnamefont {Werner}},\ }\bibfield  {title} {\bibinfo
			{title} {Dynamical mean field solution of the bose-hubbard model},\ }\href
		{https://doi.org/10.1103/PhysRevLett.105.096402} {\bibfield  {journal}
			{\bibinfo  {journal} {Phys. Rev. Lett.}\ }\textbf {\bibinfo {volume} {105}},\
			\bibinfo {pages} {096402} (\bibinfo {year} {2010})}\BibitemShut {NoStop}%
		\bibitem [{\citenamefont {Anders}\ \emph {et~al.}(2011)\citenamefont {Anders},
			\citenamefont {Gull}, \citenamefont {Pollet}, \citenamefont {Troyer},\ and\
			\citenamefont {Werner}}]{Anders2011}%
		\BibitemOpen
		\bibfield  {author} {\bibinfo {author} {\bibfnamefont {P.}~\bibnamefont
				{Anders}}, \bibinfo {author} {\bibfnamefont {E.}~\bibnamefont {Gull}},
			\bibinfo {author} {\bibfnamefont {L.}~\bibnamefont {Pollet}}, \bibinfo
			{author} {\bibfnamefont {M.}~\bibnamefont {Troyer}},\ and\ \bibinfo {author}
			{\bibfnamefont {P.}~\bibnamefont {Werner}},\ }\bibfield  {title} {\bibinfo
			{title} {Dynamical mean-field theory for bosons},\ }\href
		{https://doi.org/10.1088/1367-2630/13/7/075013} {\bibfield  {journal}
			{\bibinfo  {journal} {New Journal of Physics}\ }\textbf {\bibinfo {volume}
				{13}},\ \bibinfo {pages} {075013} (\bibinfo {year} {2011})}\BibitemShut
		{NoStop}%
		\bibitem [{\citenamefont {Akerlund}\ \emph {et~al.}(2013)\citenamefont
			{Akerlund}, \citenamefont {de~Forcrand}, \citenamefont {Georges},\ and\
			\citenamefont {Werner}}]{Akerlund2013}%
		\BibitemOpen
		\bibfield  {author} {\bibinfo {author} {\bibfnamefont {O.}~\bibnamefont
				{Akerlund}}, \bibinfo {author} {\bibfnamefont {P.}~\bibnamefont
				{de~Forcrand}}, \bibinfo {author} {\bibfnamefont {A.}~\bibnamefont
				{Georges}},\ and\ \bibinfo {author} {\bibfnamefont {P.}~\bibnamefont
				{Werner}},\ }\bibfield  {title} {\bibinfo {title} {Dynamical mean field
				approximation applied to quantum field theory},\ }\href
		{https://doi.org/10.1103/PhysRevD.88.125006} {\bibfield  {journal} {\bibinfo
				{journal} {Phys. Rev. D}\ }\textbf {\bibinfo {volume} {88}},\ \bibinfo
			{pages} {125006} (\bibinfo {year} {2013})}\BibitemShut {NoStop}%
		\bibitem [{\citenamefont {Akerlund}\ \emph {et~al.}(2014)\citenamefont
			{Akerlund}, \citenamefont {de~Forcrand}, \citenamefont {Georges},\ and\
			\citenamefont {Werner}}]{Akerlund2014}%
		\BibitemOpen
		\bibfield  {author} {\bibinfo {author} {\bibfnamefont {O.}~\bibnamefont
				{Akerlund}}, \bibinfo {author} {\bibfnamefont {P.}~\bibnamefont
				{de~Forcrand}}, \bibinfo {author} {\bibfnamefont {A.}~\bibnamefont
				{Georges}},\ and\ \bibinfo {author} {\bibfnamefont {P.}~\bibnamefont
				{Werner}},\ }\bibfield  {title} {\bibinfo {title} {Extended mean field study
				of complex ${\ensuremath{\varphi}}^{4}$-theory at finite density and
				temperature},\ }\href {https://doi.org/10.1103/PhysRevD.90.065008} {\bibfield
			{journal} {\bibinfo  {journal} {Phys. Rev. D}\ }\textbf {\bibinfo {volume}
				{90}},\ \bibinfo {pages} {065008} (\bibinfo {year} {2014})}\BibitemShut
		{NoStop}%
		\bibitem [{\citenamefont {Kim}\ \emph {et~al.}(2021)\citenamefont {Kim},
			\citenamefont {Lenk}, \citenamefont {Li}, \citenamefont {Werner},\ and\
			\citenamefont {Eckstein}}]{Aaram2021}%
		\BibitemOpen
		\bibfield  {author} {\bibinfo {author} {\bibfnamefont {A.~J.}\ \bibnamefont
				{Kim}}, \bibinfo {author} {\bibfnamefont {K.}~\bibnamefont {Lenk}}, \bibinfo
			{author} {\bibfnamefont {J.}~\bibnamefont {Li}}, \bibinfo {author}
			{\bibfnamefont {P.}~\bibnamefont {Werner}},\ and\ \bibinfo {author}
			{\bibfnamefont {M.}~\bibnamefont {Eckstein}},\ }\href
		{https://doi.org/10.48550/ARXIV.2112.15549} {\bibinfo {title} {Vertex-based
				diagrammatic treatment of light-matter-coupled systems}} (\bibinfo {year}
		{2021})\BibitemShut {NoStop}%
		\bibitem [{\citenamefont {Kim}\ \emph {et~al.}(2022)\citenamefont {Kim},
			\citenamefont {Li}, \citenamefont {Eckstein},\ and\ \citenamefont
			{Werner}}]{Aaram2022}%
		\BibitemOpen
		\bibfield  {author} {\bibinfo {author} {\bibfnamefont {A.~J.}\ \bibnamefont
				{Kim}}, \bibinfo {author} {\bibfnamefont {J.}~\bibnamefont {Li}}, \bibinfo
			{author} {\bibfnamefont {M.}~\bibnamefont {Eckstein}},\ and\ \bibinfo
			{author} {\bibfnamefont {P.}~\bibnamefont {Werner}},\ }\href
		{https://doi.org/10.48550/ARXIV.2204.13562} {\bibinfo {title}
			{Pseudo-particle vertex solver for quantum impurity models}} (\bibinfo {year}
		{2022})\BibitemShut {NoStop}%
		\bibitem [{\citenamefont {Ashida}\ \emph {et~al.}(2021)\citenamefont {Ashida},
			\citenamefont {Yokota}, \citenamefont {Imamoglu},\ and\ \citenamefont
			{Demler}}]{Ashida2021b}%
		\BibitemOpen
		\bibfield  {author} {\bibinfo {author} {\bibfnamefont {Y.}~\bibnamefont
				{Ashida}}, \bibinfo {author} {\bibfnamefont {T.}~\bibnamefont {Yokota}},
			\bibinfo {author} {\bibfnamefont {A.}~\bibnamefont {Imamoglu}},\ and\
			\bibinfo {author} {\bibfnamefont {E.}~\bibnamefont {Demler}},\ }\href
		{https://doi.org/10.48550/ARXIV.2105.08833} {\bibinfo {title}
			{Nonperturbative waveguide quantum electrodynamics}} (\bibinfo {year}
		{2021})\BibitemShut {NoStop}%
		\bibitem [{\citenamefont {Economou}(1969)}]{Economou69}%
		\BibitemOpen
		\bibfield  {author} {\bibinfo {author} {\bibfnamefont {E.~N.}\ \bibnamefont
				{Economou}},\ }\href {https://doi.org/10.1103/PhysRev.182.539} {\bibfield
			{journal} {\bibinfo  {journal} {Phys. Rev.}\ }\textbf {\bibinfo {volume}
				{182}},\ \bibinfo {pages} {539} (\bibinfo {year} {1969})}\BibitemShut
		{NoStop}%
		\bibitem [{\citenamefont {Li}\ \emph {et~al.}(2020)\citenamefont {Li},
			\citenamefont {Golez}, \citenamefont {Mazza}, \citenamefont {Millis},
			\citenamefont {Georges},\ and\ \citenamefont {Eckstein}}]{Li2020}%
		\BibitemOpen
		\bibfield  {author} {\bibinfo {author} {\bibfnamefont {J.}~\bibnamefont
				{Li}}, \bibinfo {author} {\bibfnamefont {D.}~\bibnamefont {Golez}}, \bibinfo
			{author} {\bibfnamefont {G.}~\bibnamefont {Mazza}}, \bibinfo {author}
			{\bibfnamefont {A.~J.}\ \bibnamefont {Millis}}, \bibinfo {author}
			{\bibfnamefont {A.}~\bibnamefont {Georges}},\ and\ \bibinfo {author}
			{\bibfnamefont {M.}~\bibnamefont {Eckstein}},\ }\href
		{https://doi.org/10.1103/PhysRevB.101.205140} {\bibfield  {journal} {\bibinfo
				{journal} {Phys. Rev. B}\ }\textbf {\bibinfo {volume} {101}},\ \bibinfo
			{pages} {205140} (\bibinfo {year} {2020})}\BibitemShut {NoStop}%
		\bibitem [{\citenamefont {Georges}\ \emph
			{et~al.}(1996{\natexlab{b}})\citenamefont {Georges}, \citenamefont {Kotliar},
			\citenamefont {Krauth},\ and\ \citenamefont {Rozenberg}}]{Georges96}%
		\BibitemOpen
		\bibfield  {author} {\bibinfo {author} {\bibfnamefont {A.}~\bibnamefont
				{Georges}}, \bibinfo {author} {\bibfnamefont {G.}~\bibnamefont {Kotliar}},
			\bibinfo {author} {\bibfnamefont {W.}~\bibnamefont {Krauth}},\ and\ \bibinfo
			{author} {\bibfnamefont {M.~J.}\ \bibnamefont {Rozenberg}},\ }\bibfield
		{title} {\bibinfo {title} {Dynamical mean-field theory of strongly correlated
				fermion systems and the limit of infinite dimensions},\ }\href
		{https://doi.org/10.1103/RevModPhys.68.13} {\bibfield  {journal} {\bibinfo
				{journal} {Rev. Mod. Phys.}\ }\textbf {\bibinfo {volume} {68}},\ \bibinfo
			{pages} {13} (\bibinfo {year} {1996}{\natexlab{b}})}\BibitemShut {NoStop}%
		\bibitem [{\citenamefont {Ayral}\ \emph {et~al.}(2013)\citenamefont {Ayral},
			\citenamefont {Biermann},\ and\ \citenamefont {Werner}}]{Ayral2013}%
		\BibitemOpen
		\bibfield  {author} {\bibinfo {author} {\bibfnamefont {T.}~\bibnamefont
				{Ayral}}, \bibinfo {author} {\bibfnamefont {S.}~\bibnamefont {Biermann}},\
			and\ \bibinfo {author} {\bibfnamefont {P.}~\bibnamefont {Werner}},\
		}\bibfield  {title} {\bibinfo {title} {Screening and nonlocal correlations in
				the extended hubbard model from self-consistent combined $gw$ and dynamical
				mean field theory},\ }\href {https://doi.org/10.1103/PhysRevB.87.125149}
		{\bibfield  {journal} {\bibinfo  {journal} {Phys. Rev. B}\ }\textbf {\bibinfo
				{volume} {87}},\ \bibinfo {pages} {125149} (\bibinfo {year}
			{2013})}\BibitemShut {NoStop}%
		\bibitem [{\citenamefont {Werner}\ \emph {et~al.}(2006)\citenamefont {Werner},
			\citenamefont {Comanac}, \citenamefont {de' Medici}, \citenamefont {Troyer},\
			and\ \citenamefont {Millis}}]{Werner2006}%
		\BibitemOpen
		\bibfield  {author} {\bibinfo {author} {\bibfnamefont {P.}~\bibnamefont
				{Werner}}, \bibinfo {author} {\bibfnamefont {A.}~\bibnamefont {Comanac}},
			\bibinfo {author} {\bibfnamefont {L.}~\bibnamefont {de' Medici}}, \bibinfo
			{author} {\bibfnamefont {M.}~\bibnamefont {Troyer}},\ and\ \bibinfo {author}
			{\bibfnamefont {A.~J.}\ \bibnamefont {Millis}},\ }\bibfield  {title}
		{\bibinfo {title} {Continuous-time solver for quantum impurity models},\
		}\href {https://doi.org/10.1103/PhysRevLett.97.076405} {\bibfield  {journal}
			{\bibinfo  {journal} {Phys. Rev. Lett.}\ }\textbf {\bibinfo {volume} {97}},\
			\bibinfo {pages} {076405} (\bibinfo {year} {2006})}\BibitemShut {NoStop}%
		\bibitem [{\citenamefont {Weber}(2022)}]{Weber2022}%
		\BibitemOpen
		\bibfield  {author} {\bibinfo {author} {\bibfnamefont {M.}~\bibnamefont
				{Weber}},\ }\bibfield  {title} {\bibinfo {title} {Quantum monte carlo
				simulation of spin-boson models using wormhole updates},\ }\href
		{https://doi.org/10.1103/PhysRevB.105.165129} {\bibfield  {journal} {\bibinfo
				{journal} {Phys. Rev. B}\ }\textbf {\bibinfo {volume} {105}},\ \bibinfo
			{pages} {165129} (\bibinfo {year} {2022})}\BibitemShut {NoStop}%
		\bibitem [{\citenamefont {Jeckelmann}\ \emph {et~al.}(1999)\citenamefont
			{Jeckelmann}, \citenamefont {Zhang},\ and\ \citenamefont
			{White}}]{Jeckelmann1999}%
		\BibitemOpen
		\bibfield  {author} {\bibinfo {author} {\bibfnamefont {E.}~\bibnamefont
				{Jeckelmann}}, \bibinfo {author} {\bibfnamefont {C.}~\bibnamefont {Zhang}},\
			and\ \bibinfo {author} {\bibfnamefont {S.~R.}\ \bibnamefont {White}},\
		}\bibfield  {title} {\bibinfo {title} {Metal-insulator transition in the
				one-dimensional holstein model at half filling},\ }\href
		{https://doi.org/10.1103/PhysRevB.60.7950} {\bibfield  {journal} {\bibinfo
				{journal} {Phys. Rev. B}\ }\textbf {\bibinfo {volume} {60}},\ \bibinfo
			{pages} {7950} (\bibinfo {year} {1999})}\BibitemShut {NoStop}%
		\bibitem [{\citenamefont {Jansen}\ \emph {et~al.}(2021)\citenamefont {Jansen},
			\citenamefont {Jooss},\ and\ \citenamefont {Heidrich-Meisner}}]{Jansen2021}%
		\BibitemOpen
		\bibfield  {author} {\bibinfo {author} {\bibfnamefont {D.}~\bibnamefont
				{Jansen}}, \bibinfo {author} {\bibfnamefont {C.}~\bibnamefont {Jooss}},\ and\
			\bibinfo {author} {\bibfnamefont {F.}~\bibnamefont {Heidrich-Meisner}},\
		}\bibfield  {title} {\bibinfo {title} {Charge density wave breakdown in a
				heterostructure with electron-phonon coupling},\ }\href
		{https://doi.org/10.1103/PhysRevB.104.195116} {\bibfield  {journal} {\bibinfo
				{journal} {Phys. Rev. B}\ }\textbf {\bibinfo {volume} {104}},\ \bibinfo
			{pages} {195116} (\bibinfo {year} {2021})}\BibitemShut {NoStop}%
		\bibitem [{\citenamefont {Biermann}\ \emph {et~al.}(2003)\citenamefont
			{Biermann}, \citenamefont {Aryasetiawan},\ and\ \citenamefont
			{Georges}}]{Biermann2003}%
		\BibitemOpen
		\bibfield  {author} {\bibinfo {author} {\bibfnamefont {S.}~\bibnamefont
				{Biermann}}, \bibinfo {author} {\bibfnamefont {F.}~\bibnamefont
				{Aryasetiawan}},\ and\ \bibinfo {author} {\bibfnamefont {A.}~\bibnamefont
				{Georges}},\ }\bibfield  {title} {\bibinfo {title} {First-principles approach
				to the electronic structure of strongly correlated systems: Combining the
				$gw$ approximation and dynamical mean-field theory},\ }\href
		{https://doi.org/10.1103/PhysRevLett.90.086402} {\bibfield  {journal}
			{\bibinfo  {journal} {Phys. Rev. Lett.}\ }\textbf {\bibinfo {volume} {90}},\
			\bibinfo {pages} {086402} (\bibinfo {year} {2003})}\BibitemShut {NoStop}%
		\bibitem [{\citenamefont {Steinke}\ \emph {et~al.}(2020)\citenamefont
			{Steinke}, \citenamefont {Wehling},\ and\ \citenamefont
			{R\"osner}}]{Steinke2020}%
		\BibitemOpen
		\bibfield  {author} {\bibinfo {author} {\bibfnamefont {C.}~\bibnamefont
				{Steinke}}, \bibinfo {author} {\bibfnamefont {T.~O.}\ \bibnamefont
				{Wehling}},\ and\ \bibinfo {author} {\bibfnamefont {M.}~\bibnamefont
				{R\"osner}},\ }\bibfield  {title} {\bibinfo {title} {Coulomb-engineered
				heterojunctions and dynamical screening in transition metal dichalcogenide
				monolayers},\ }\href {https://doi.org/10.1103/PhysRevB.102.115111} {\bibfield
			{journal} {\bibinfo  {journal} {Phys. Rev. B}\ }\textbf {\bibinfo {volume}
				{102}},\ \bibinfo {pages} {115111} (\bibinfo {year} {2020})}\BibitemShut
		{NoStop}%
	\end{thebibliography}
	%
	\newpage
	\onecolumngrid
	\appendix
	\section{Quantization of the SPP mode}
	\label{sec:quant_SPP}
	To quantize the SPP mode, we start from a classical description and solve the macroscopic Maxwell's equations for the metal-dielectric interface, closely following Ref.~\cite{Economou69}. We assume that the relative permeability is given by $\mu=1$ in both the dielectric and the metal region. Since there are no free charges and currents, Maxwell's equations are given by 
	\begin{align}
	\bm{\nabla}\cdot \bm D &= 0, \\
	\bm{\nabla}\cdot \bm H &= 0, \\
	\bm{\nabla}\times\bm E &= -\mu_0\frac{\partial \bm H}{\partial t}, \label{equ:curlE}\\
	\bm{\nabla}\times\bm H &= \epsilon_0\epsilon_r\frac{\partial \bm E}{\partial t}.\label{equ:curlH}
	\end{align}
	Without loss of generality, we first solve the equations for an SPP mode travelling in the $z$-direction using the ansatz $\bm{F}=\bm{\mathcal{F}}(x)e^{i(q z - \omega t)}$ for all fields. Later, we can generalize the solution and sum all possible directions of propagation. Moreover, we restrict our considerations to the transverse magnetic (TM) mode with $E_y=H_x=H_z=0$.
	With this, we obtain
	\begin{align}
	i\bm{q}\mathcal{H}_y(x)=&i\omega\epsilon_0\epsilon(x,\omega)\mathcal{E}_x(x) \label{equ:curlHx}\\
	\partial_x\mathcal{H}_y(x)=&-i\omega\epsilon_0\epsilon(x,\omega)\mathcal{E}_z(x) \label{equ:curlHz}
	\end{align}
	from the $x$- and the $z$-component of Eq.~\eqref{equ:curlH}, as well as
	\begin{equation}
		i\bm{q}\mathcal{E}_x(x)-\partial_x\mathcal{E}_z(x)=\mu_0i\omega\mathcal{H}_y(x) \label{equ:curlEy}
	\end{equation}
	using the $y$-component of Eq.~\eqref{equ:curlE}. Combining \eqref{equ:curlHx} and \eqref{equ:curlHz} we get
	\begin{equation}
		\mathcal{E}_z(x)=-\frac{i}{q}\partial_x\mathcal{E}_x(x).
		\label{equ:Ez}
	\end{equation}
	Furthermore, we substitute $\mathcal{H}_y(x)$ and $\mathcal{E}_z(x)$ in Eq.~\eqref{equ:curlEy} using Eq.~\eqref{equ:curlHx} and \eqref{equ:Ez}, which yields
	\begin{equation}
		0=\partial_x^2\mathcal{E}_x(x)-Q^2\mathcal{E}_x(x) \label{equ:wave_eq}
	\end{equation}
	with 
	\begin{equation}
		Q(x) = \sqrt{q ^2-\left(\frac{\omega}{c}\right)^2\epsilon(x,\omega)}.
	\end{equation}
	Equation~\eqref{equ:wave_eq} is solved by 
	\begin{equation}
	\mathcal{E}_x(x)=\begin{cases}
	A_{\mathrm{m}}e^{Q_\mathrm{m} x} & ,x<0 \\
	A_{\mathrm{d}}e^{Q_\mathrm{d} x} & ,x>0
	\end{cases},
	\end{equation}
	where $\epsilon_{\mathrm{m}}$ and $\epsilon_{\mathrm{d}}$ refer to the dielectric constant in the metallic ($x<0$) and the dielectric ($x>0$) region, respectively. 
	
	Using the continuity of the tangential components of $\mathbf{E}$ and $\mathbf{H}$ at the interface, we obtain the relation
	\begin{equation}
	A_{\mathrm{m}}=\frac{\epsilon_{\mathrm{d}}}{\epsilon_{\mathrm{m}}}A_{\mathrm{d}}
	\end{equation}
	for the field amplitudes. This can be seen from Eq.~\eqref{equ:curlHx}.
	Moreover, together with Eq.~\eqref{equ:Ez}, this yields the dispersion relation 
	\begin{equation}
	\frac{Q_{\mathrm{m}}}{\epsilon_{\mathrm{m}}}=-\frac{Q_{\mathrm{d}}}{\epsilon_{\mathrm{d}}},
	\end{equation}
	which can be solved for $q$ and reads
	\begin{equation}
		q=\frac{\omega}{c}\sqrt{\frac{\epsilon_{\mathrm{m}}\epsilon_{\mathrm{d}}}{\epsilon_{\mathrm{m}}+\epsilon_{\mathrm{d}}}}.
		\label{equ:general_disp}
	\end{equation}
	In the following, we assume a frequency independent dielectric permittivity $\epsilon_{\mathrm{d}}$ and a simple Drude response for the metal, i.e.
	\begin{equation}
	\epsilon_{\mathrm{m}}(\omega)=1-\left(\frac{\omega_{\mathrm p}}{\omega}\right)^2 ,
	\end{equation}
	where $\omega_{\mathrm p}$ denotes the plasma frequency.
	Even though, with this, Eq.~\eqref{equ:general_disp} cannot be solved analytically for the frequency, we may investigate the asymptotic behavior of $\omega(q)$. Taylor-expanding $1/\epsilon_{\rm m}$ in powers of $\omega$ shows that for small $\omega$ this term vanishes up to second order; therefore, it may be ignored for $\omega\ll0$ such that $\omega\approx qc/\sqrt{\epsilon_{\mathrm{d}}}$. On the other hand, for real valued $q$ (i.e. propagating modes) the argument of the square root has to be positive. From this, we obtain the condition $\omega<\omega_{\rm p}/\sqrt{\epsilon_{\mathrm{d}}+1}$. This is exactly where Eq.~\eqref{equ:general_disp} diverges; thus the frequency saturates at this value as $q\rightarrow\infty$. Figure~\ref{fig:disp} in the main text shows the dispersion for three different permittivities $\epsilon_{\mathrm{d}}$. The dotted and dashed lines indicate the asymptotic values for $\omega\ll1$ and $\omega\rightarrow\infty$, respectively. For the DMFT calculations, we restrict our considerations to the simplest case, where $\epsilon_{\mathrm{d}}=1$. With this, the dispersion relation can even be solved analytically for the frequency and has only one physical solution, which reads
	\begin{equation}
	\omega_{q}=\sqrt{\frac{\omega_{\mathrm p}^2}{2}+q^2c^2-\sqrt{\frac{\omega_{\mathrm p}^4}{4}+q^4c^4}}.
	\end{equation}
	
	To quantize the theory, we define the bosonic photon annihilation (creation) operators $\hat{a}_{\bm{q}}^{(\dagger)}$. With this the electric field operator can be expanded as follows:
	\begin{equation}
	\hat{\bm E}(\bm r)=\sum_{\bm q}\sqrt{\frac{\omega_{\bm q}}{2\epsilon_0\epsilon(x,\omega_{\bm q})Na^3}}[\bm{u}_{\bm q}(x)e^{i\bm{q}\cdot\bm{\rho}}\hat{a}_{\bm q}+h.c.],
	\end{equation}
	where $a$ denotes the lattice constant and $N$ represents the number of emitters.
	Here, we have introduced the two-dimensional vectors $\bm{\rho}=(y,z)^T$ and $\bm q=(q_y,q_z)^T=q(\sin(\varphi),\cos(\varphi))^T$. The latter includes all possible TM SPP-modes travelling along the interface (not only those propagating in the $z$-direction), where $\varphi\in[0,2\pi)$ denotes the angle to the $z$-axis and $q>0$.
	Since (for $\hbar=1$) the Hamiltonian of the free electromagnetic field should be given by 
	\begin{equation}
	\hat{H}_{\mathrm{field}}=\sum_{\bm q}\omega_{\bm q} \hat{a}_{\bm q}^\dagger \hat{a}_{\bm q},
	\end{equation}
	the mode functions
	\begin{equation}
	\bm{u}_{\bm q}(x)=\mathcal{N}_{\bm q}\begin{cases}
	e^{Q_{\mathrm{m}}x}\begin{pmatrix}
	1 \\
	-i(Q_{\mathrm{m}}/q)\sin(\varphi) \\
	-i(Q_{\mathrm{m}}/q)\cos(\varphi)
	\end{pmatrix}, & x<0\\
	e^{-Q_{\mathrm{d}}x}\begin{pmatrix}
	1 \\
	i(Q_{\mathrm{d}}/q)\sin(\varphi) \\
	i(Q_{\mathrm{d}}/q)\cos(\varphi)
	\end{pmatrix}, & x>0
	\end{cases}
	\end{equation}
	must satisfy the condition $\int_V d^3r|\bm u_{\bm q}(x)|^2=Na^3$. Thus, the normalization factor is given by
	\begin{equation}
	\mathcal{N}_{\bm q}=\sqrt{a}\left\{\frac{1}{2Q_{\mathrm{m}}}\left[1+\left(\frac{Q_{\mathrm{m}}}{q}\right)^2\right]+\frac{1}{2Q_{\mathrm{d}}}\left[1+\left(\frac{Q_{\mathrm{d}}}{q}\right)^2\right]\right\}^{-\frac{1}{2}}.
	\end{equation}

		Note that this factor would vanish if we had considered the vacuum case, where the electromagnetic field can be expanded in simple plane waves, i.e., ${\bm u}_{\bm  q}( x)\sim e^{iq_x x}$. Therefore, the exponential decay and the resulting confinement of the SPP mode to the dielectric-metal interface leads to an enhancement of the light-matter interaction as compared to free space.
	
	\section{Photon-induced interaction}
	\label{sec:ind_int_app}
	The full imaginary-time action can be split into four parts
	\begin{equation}
	S=S_{\rm{mat}}+S_{\rm{PP}}+S_{\rm{EP}}+S_{\rm{field}}.
	\end{equation}
	In this representation, the photonic degrees of freedom are represented by bosonic variables $a_{\bm q}(\tau)$ and $\bar{a}_{\bm q}(\tau)$, such that the action of the free electromagnetic field reads
	\begin{equation}
		S_{\rm{field}}=\int\limits_0^{\beta}d\tau \bar{a}_{\bm q}(\tau)\left[\partial_{\tau}+\omega_{\bm q}\right]a_{\bm{q}}(\tau).
	\end{equation}
	and the light-matter coupling term $S_{\rm{EP}}$ is given by
	\begin{equation}
		S_{\rm{EP}}=\int\limits_0^{\beta}d\tau \sum_{r,\bm q}\sqrt{\frac{\omega_{\bm q}}{2N}}\left[g_{\bm q}e^{i\bm{q}\cdot\bm{R}_{r}}a_{\bm q}(\tau)+c.c.\right]\sigma_r^1(\tau).
	\end{equation}
	To obtain an effective description for the material, we trace out the photon fields from the full action. This allows us to incorporate the effect of the light-matter coupling in a single induced interaction term $S_{\rm{ind}}$ that satisfies the relation 
	\begin{equation}
	e^{-S_{\rm{ind}}=}e^{-S_{\rm{PP}}}\int\mathcal{D}[\bar{a},a]e^{-(S_{\rm{EP}}+S_{\rm{field}})}.
	\end{equation}
	The action is only quadratic in the bosonic fields $a_{\bm q}(\tau)$ and $\bar{a}_{\bm q}(\tau)$; therefore, the second factor is just a simple Gaussian path integral with quadratic extension, which can be solved analytically. From this, we obtain a retarded dipole-dipole interaction
	\begin{equation}
	S_{\rm{ind}}=-\frac{1}{2}\int\limits_0^\beta d\tau\int\limits_0^\beta d\tau'\sum_{r,r'}\sigma_r^1(\tau)W^{\rm ind}_{r,r'}(\tau-\tau')\sigma_{r'}^1(\tau'),
	\end{equation}
	where the interaction vertex reads	
	\begin{equation}
	W^{\rm{ind}}_{r,r'}(\tau)=-\sum_{\bm q}\frac{|g_{\bm q}|^2}{N}e^{-i\bm q\cdot(\bm R_r-\bm R_{r'})}\left[1+\omega_{\bm q}D^{0}_{\bm q}(\tau)\right].
	\end{equation}
	Here, $D_{\bm{q}}^0(\tau)$ denotes the free photon propagator and is given by
	\begin{equation}
		D_{\bm q}^0(\tau)=-\langle a_{\bm q}(\tau)\bar{a}_{\bm q}(0) \rangle_{S_{\rm{field}}}=-\frac{e^{-\tau\omega_{\bm q}}}{1-e^{-\beta\omega_{\bm q}}},
	\end{equation}
	or
	\begin{equation}
	D^{0}_{\bm q}(i\nu_m)=\frac{1}{i\nu_m-\omega_{\bm q}}
	\end{equation}
	in the Matsubara representation. Since the induced interaction $W^{\rm{ind}}_{r,r'}(i\nu_n)$ only depends on the distance $\bm{R}_r-\bm{R}_{r'}$, we may perform the lattice Fourier transform
	\begin{equation}
	f_{\bm k}=\sum_{r} f_re^{i\bm k \cdot \bm R _r},
	\label{equ:latticeFT}
	\end{equation}
	which yields
	\begin{equation}
	W_{\bm k}(i\nu_n)=-\sum_{\bm{G}\in\mathcal{L}_{\rm R}}|g_{\bm {k}+\bm{G}}|^2\frac{\nu_n^2}{\nu_n^2+\omega_{\bm{k}+\bm{G}}^2},
	\end{equation}
	where $\mathcal{L}_{\rm R}$ denotes the reciprocal lattice.
	However, as shown in Fig.~\ref{fig:gq2} in the main text, the strength of the coupling $|g_{\bm q}|^2$ decreases as $q$ increases. In particular, we can assume that
	$|g_{\bm {k}+\bm{G}}|^2 \rightarrow 0$ for $\bm{G}\neq\bm{0}$. With this the result simplifies to
	\begin{equation}
	W_{\bm k}(i\nu_n)=-|g_{\bm k}|^2\frac{\nu_n^2}{\nu_n^2+\omega_{\bm k}^2}.
	\end{equation}
	
	\section{Details on the mean-field approximation}
	\label{sec:MF_app}
	\subsection{Cancellation of the light-matter interaction terms}
	\label{sec:MF_app_cancel}
	In this section we show how the effect of the light-matter interaction on the static mean-field result is cancelled if all relevant terms are taken into account. We start by decoupling the dipole-dipole interactions using the substitution  
	$\hat{\sigma}_{r}^x\hat{\sigma}_{r'}^1\rightarrow\langle\sigma^1\rangle\hat{\sigma}_{r'}^1+\hat{\sigma}_{r}^x\langle\sigma^1\rangle$.
	This means that quantum fluctuations of the dipolar moments around the mean-field expectation value are neglected beyond the first order. With this, the nearest neighbor interaction can be rewritten as 
	\begin{equation}
		\hat{H}_{\rm{nn}}^{\rm{mf}}=\alpha\langle \sigma^1\rangle\sum_{r}\hat{\sigma}^x_{r}
	\end{equation}
	and the quadratic term in the light-matter Hamiltonian becomes
	\begin{equation}
		\hat{H}_{\rm{PP}}^{\rm{mf}}=\sum_{r,r'}\sum_{\bm q}\frac{|g_{\bm q}|^2}{N}e^{-i\bm q\cdot(\bm{R}_{r}-\bm{R}_{r'})}\langle \sigma^1\rangle\hat{\sigma}_{r'}^1.
	\end{equation}
	Moreover, we perform the replacement
	\begin{equation}
		\left[g_{\bm q}e^{i\bm{q}\cdot\bm{R}_{r}}\hat{a}_{\bm q}+h.c.\right]\hat{\sigma}_r^1
		\rightarrow
		\left[g_{\bm q}e^{i\bm{q}\cdot\bm{R}_{r}}\langle \hat{a}_{\bm q}\rangle+c.c.\right]\hat{\sigma}_r^1
		+
		\left[g_{\bm q}e^{i\bm{q}\cdot\bm{R}_{r}}\hat{a}_{\bm q}+h.c.\right]\langle \sigma^1\rangle
	\end{equation}
	to decouple light and matter, such that the linear coupling term can be split into two terms
	\begin{equation}
	\hat{H}_{\rm EP}\rightarrow\hat{H}_{\rm EP}^{\rm mf, mat}+\hat{H}_{\rm EP}^{\rm mf, field},
	\end{equation}
	where
	\begin{equation}
		\hat{H}_{\rm EP}^{\rm mf, mat}=
		\sum_{r,\bm q}\sqrt{\frac{\omega_{\bm q}}{2N}}\left[g_{\bm q}e^{i\bm{q}\cdot\bm{R}_{r}}\langle\hat{a}_{\bm q}\rangle+h.c.\right]\hat{\sigma}_r^1
		\label{equ:H_EPmfmat}
	\end{equation}
	only depends on the matter operators and the light enters via expectation values. Similarly,
	\begin{equation}
		\hat{H}_{\rm EP}^{\rm mf, field}=\sum_{r,\bm q}\sqrt{\frac{\omega_{\bm q}}{2N}}\left[g_{\bm q}e^{i\bm{q}\cdot\bm{R}_{r}}\hat{a}_{\bm q}+h.c.\right]\langle\sigma^1\rangle
	\end{equation}
	only contains photonic operators and expectation values of matter operators. Likewise, we may separate the photonic and the matter operators in the total Hamiltonian
	\begin{equation}
		\hat{H}\rightarrow\hat{H}_{\rm field}^{\rm mf}+\hat{H}_{\rm mat}^{\rm mf}
	\end{equation}
	with
	\begin{equation}
		\hat{H}_{\rm field}^{\rm mf}
		=\hat{H}_{\rm field}
		+\hat{H}_{\rm EP}^{\rm mf, field}
	\end{equation}
	and
	\begin{equation}
		\hat{H}_{\rm mat}^{\rm mf}
		=\hat{H}_0
		+\hat{H}_{\rm nn}^{\rm mf}
		+\hat{H}_{\rm PP}^{\rm mf}
		+\hat{H}_{\rm EP}^{\rm mf,mat}.
	\end{equation}
	Let us now consider the semiclassical equation of motion for the photon annihilation operator
	\begin{equation}
		\langle \dot{\hat{a}}_{\bm q} \rangle
		=i\left\langle\left[\hat{H}_{\rm field}^{\rm mf},\hat{a}_{\bm q}\right]\right\rangle
		=-i\omega_{\rm q}\langle \hat{a}_{\bm q}\rangle-i\sum_{r}\sqrt{\frac{\omega_{\bm q}}{2N}}\bar{g}_{\bm q}e^{-i\bm{q}\cdot\bm{R}_r}\sigma^1.
	\end{equation}
	The corresponding expression for the creation operator is obtained by taking the complex conjugate. For the static solution, the time derivatives vanish, which yields
	\begin{align}
		\langle\hat{a}_{\bm q}\rangle&=-\sum_{r}\sqrt{\frac{1}{2N\omega_{\bm q}}}\bar{g}_{\bm q}e^{-i\bm{q}\cdot\bm{R}_r}\sigma^1\\
		\langle\hat{a}_{\bm q}^\dagger\rangle&=-\sum_{r}\sqrt{\frac{1}{2N\omega_{\bm q}}}g_{\bm q}e^{i\bm{q}\cdot\bm{R}_r}\sigma^1.
	\end{align}
	With this, we can eliminate the expectation values $\langle\hat{a}_{\bm q}\rangle$ and $\langle\hat{a}_{\bm q}^\dagger\rangle$ in Eq.~\eqref{equ:H_EPmfmat} and obtain
	\begin{equation}
		\hat{H}_{\rm EP}^{\rm mat}=-\sum_{r,r'}\sum_{\bm q}\frac{|g_{\bm q}|^2}{N}e^{-i\bm q\cdot(\bm{R}_{r}-\bm{R}_{r'})}\langle \sigma^1\rangle\hat{\sigma}_{r'}^1=-\hat{H}_{PP}^{\rm mf}.
	\end{equation}
	This expression proves that the linear coupling term between the electric displacement field and the polarization exactly cancels the quadratic term $\hat{H}_{PP}^{\rm mf}$. As a result, within the mean-field approximation, all contributions due to the light-matter interaction disappear and the system is described by the uncoupled Hamiltonian
	\begin{equation}
		\hat{H}_{\rm mat}^{\rm mf}
		=\hat{H}_0
		+\hat{H}_{\rm nn}^{\rm mf}.
		\label{equ:H_mf_final}
	\end{equation}
	
	\subsection{Self-consistent mean-field equation}
	\label{sec:MF_app_selfcons}
	Starting from the mean-field Hamiltonian \eqref{equ:H_mf_final}, we derive a self-consistent equation for the order parameter that allows to determine critical values for the temperature and the dipole-dipole interaction. For that purpose, we introduce an additional external static field $f$ that couples to the total polarization of the material and define
	\begin{equation}
		\hat{H}^{\rm mf}[f]=\hat{H}_{\rm mat}^{\rm mf}-f\sum_{r}\hat{\sigma}_r^1=\sum_{r}\hat{H}_r^{\rm mf}[f]
	\end{equation}
	with the single-site mean-field Hamiltonian
	\begin{equation}
		\hat{H}_r^{\rm mf}[f]=\frac{\Delta}{2}\hat{\sigma}_{r}^z-\alpha\langle\sigma^1\rangle\hat{\sigma}_r^1-f\hat{\sigma}_r^1.
	\end{equation}
	The latter can be diagonalized, which yields the eigenvalues
	\begin{equation}
		E_{\pm}[f]=\pm\sqrt{(\Delta/2)^2+(\alpha\langle\sigma^1\rangle-f)^2}.
	\end{equation}
	Let us also introduce the partition function
	\begin{equation}
		\mathcal{Z}_r[f]={\rm{tr}}\left\{e^{-\beta\hat{H}_r^{\rm mf}[f]}\right\}=e^{-\beta E_+[f]}+e^{-\beta E_-[f]}.
	\end{equation}
	Then, the order parameter can be calculated by taking the derivative
	\begin{equation}
		\langle\sigma^1\rangle=\frac{1}{\beta}\frac{\partial}{\partial f}{\rm{ln}}\mathcal{Z}_{r}[f]\bigg\vert_{f=0}. 
	\end{equation}
	From this, we obtain the self-consistent equation
	\begin{equation}
		\langle\sigma^1\rangle=\tanh\left[\beta\sqrt{(\Delta/2)^2+(\alpha\langle\sigma^1\rangle)^2}\right]\frac{\alpha}{\sqrt{(\Delta/2)^2+(\alpha\langle\sigma^1\rangle)^2}}\langle\sigma^1\rangle.
		\label{equ:sc_equ}
	\end{equation}
	It can be solved numerically using a fixed point iteration. Results are shown in the main text in Fig.~\ref{fig:mf}. In general, it always has a trivial solution $\langle\sigma^1\rangle = 0$; however, this solution is only stable in the normal paraelectric  phase. For the ferroelectric state 	$\langle\sigma^1\rangle\neq 0$, such that Eq.~\eqref{equ:sc_equ} can be rewritten as  
	\begin{equation}
	1=\tanh\left[\beta\sqrt{(\Delta/2)^2+(\alpha\langle\sigma^1\rangle)^2}\right]\frac{\alpha}{\sqrt{(\Delta/2)^2+(\alpha\langle\sigma^1\rangle)^2}}.
	\end{equation}
	For $\beta\rightarrow\infty$, the right-hand side goes to $\frac{\alpha}{\sqrt{(\Delta/2)^2+(\alpha\langle\sigma^1\rangle)^2}}$. Since $\beta=1/T$, this is the zero-temperature limit. Therefore, the ferroelectric phase only exists if $\frac{\alpha}{\sqrt{(\Delta/2)^2+(\alpha\langle\sigma^1\rangle)^2}}\ge1$. In particular this means that it is impossible to undergo a transition to the ordered phase if $\alpha\le\Delta/2$ 
	even at $T=0$. Since, in this case, the ferroelectric state is not destabilized due to classical thermal fluctuations but due to the microscopic splitting of the energy levels, this is the so-called quantum paraelectric regime.
	
	\section{Derivation of the DMFT equations}
	\subsection{Mapping of the lattice action to an impurity problem}
	\label{sec:map_to_imp}
	In this section, we map the full lattice model to a local impurity problem using the cavity method. We start from the Hubbard-Stratonovich action 
	\begin{equation}
	S_{\rm HS}=S_{0}+S_{\varphi\varphi}+S_{\varphi\sigma}
	\end{equation}
	with
	\begin{equation}
	S_{\varphi\varphi}=	\frac{1}{2}\int\limits_0^{\beta}d\tau\int\limits_0^{\beta}d\tau'\sum_{r,r'}\varphi_r(\tau)[W^{-1}]_{r,r'}(\tau-\tau')\varphi_{r'}(\tau')
	\end{equation}
	and
	\begin{equation}
	S_{\varphi\sigma}=-\sum_{r}\int\limits_0^{\beta}d\tau\varphi_r(\tau)\sigma_r^1(\tau)
	\end{equation}
	that has been derived in the main text in Sec.~\ref{sec:HS-trafo}.
	The action of the non-interacting material can be rewritten as a sum over all sites $r$
	\begin{equation}
	S_{0}=\sum_{r}S_{0}^r,
	\end{equation}
	where  $S_{0}^r$ is the action of one isolated emitter. Now we single out one site $c$ of the lattice - the so-called cavity site - and split the action into three contributions. The first one contains all on-site terms
	\begin{equation}
	S^{c}=S_{0}^c+S_{\varphi\varphi}^c+S_{\varphi\sigma}^c,
	\end{equation}
	with
	\begin{equation}
	S_{\varphi\varphi}^c=\frac{1}{2}\int\limits_0^{\beta}d\tau\int\limits_0^{\beta}d\tau'\varphi_c(\tau)[W^{-1}]_{c,c}(\tau-\tau')\varphi_{c}(\tau').
	\end{equation}
	and
	\begin{equation}
	S_{\varphi\sigma}^c=-\int\limits_0^{\beta}d\tau\varphi_c(\tau)\sigma_c^1(\tau).
	\end{equation}
	The second 
	contribution
	is the action of the system with one missing emitter, i.e. a cavity, at site $c$
	\begin{equation}
	S^{(c)}=\sum_{r\neq c}S_{0}^r+S_{\varphi\varphi}^{(c)}+S_{\varphi\sigma}^{(c)},
	\end{equation}
	where
	\begin{equation}
	S_{\varphi\varphi}^{(c)}=\frac{1}{2}\int\limits_0^{\beta}d\tau\int\limits_0^{\beta}d\tau'\sum_{r\neq c}\sum_{r'\neq c}\varphi_r(\tau)[W^{-1}]_{r,r'}(\tau-\tau')\varphi_{r'}(\tau')
	\label{equ:cav_latt_action}
	\end{equation}
	and
	\begin{equation}
	S_{\varphi\sigma}^{(c)}=-\sum_{r\neq c}\int\limits_0^{\beta}d\tau\varphi_r(\tau)\sigma_r^1(\tau).
	\end{equation}
	The third contribution describes the interaction between the cavity site and the rest of the lattice and reads
	\begin{equation}
		\Delta S=\sum_{r\neq c}\int\limits_0^{\beta}d\tau\varphi_r(\tau)t_r(\tau)
	\end{equation}
	with
	\begin{equation}
	t_r(\tau)=\int\limits_0^{\beta}d\tau'[W^{-1}]_{r,c}(\tau-\tau')\varphi_c(\tau').
	\end{equation}
	In the next step, we integrate out all matter fields $\{\xi_r,\bar{\xi}_r\}$ and auxiliary fields $\{\varphi_{r}\}$ with $r\neq c$, such that the effect of all other sites on the impurity can be described by an effective action $S_{\rm hyb}$, i.e.
	\begin{equation}
	\int\mathcal{D}[\varphi_{r\neq c}]\int\mathcal{D}[\xi_{r\neq c}\bar{\xi}_{r\neq c}]e^{-S_{HS}}=e^{-S^c}\int\mathcal{D}[\varphi_{r\neq c}]\int\mathcal{D}[\xi_{r\neq c}\bar{\xi}_{r\neq c}]e^{-(S^{(c)}+\Delta S)}=e^{-(S^c+S^{\rm hyb})}.
	\end{equation}
	The path integral can be solved formally using a cumulant expansion in $\Delta S$. With this
	\begin{equation}
	S^{\rm hyb}=-\sum_{n=1}^{\infty}\frac{(-1)^n}{n!}\sum_{r_1...r_n\neq 0}\int\limits_0^\beta d\tau_1...\int\limits_0^{\beta}d\tau_n t_{r_n}(\tau_1)...t_{r_n}(\tau_n)K_{r_1...r_n}(\tau_1...\tau_n),
	\end{equation}
	where the connected correlation functions $K_{r_1...r_n}(\tau_1...\tau_n)$ are defined as 
	\begin{equation}
	K_{r_1...r_n}(\tau_1...\tau_n)=\langle\mathcal{T}\varphi_{r_1}(\tau_1)...\varphi_{r_n}(\tau_n)\rangle_{S^{(c)}}^{\rm con}.
	\end{equation}
	The subscript $S^{(c)}$ indicates that the time-ordered expectation value is evaluated using the action \eqref{equ:cav_latt_action}.
	For our DMFT calculations, we truncate all terms beyond the second order. With this
	\begin{equation}
		S^{\rm hyb}=S_1^{\rm hyb}+S_2^{\rm hyb},
	\end{equation}
	where the first order term is given by
	\begin{equation}
	S_1^{\rm hyb}=\int\limits_0^{\beta}d\tau\varphi_c(\tau)h(\tau)
	\end{equation}
	with
	\begin{equation}
	h(\tau)=\sum_{r\neq c}\int\limits_0^{\beta}d\tau'[W^{-1}]_{c,r}(\tau-\tau')\langle\mathcal{T}\varphi_r(\tau')\rangle_{S^{(c)}},
	\end{equation}
	and the second order term reads
	\begin{equation}
	S_2^{\rm hyb}=-\int\limits_0^{\beta}d\tau\int\limits_0^{\beta}d\tau'\varphi_c(\tau)\Delta_{\rm hyb}(\tau-\tau')\varphi_c(\tau')
	\end{equation}
	with the hybridization function
	\begin{equation}
	\Delta_{\rm hyb}(\tau)=\sum_{r\neq c}\sum_{r'\neq c}\int\limits_0^{\beta}d\tau_1\int\limits_0^{\beta}d\tau_2 [W^{-1}]_{c,r}(\tau-\tau_1)U_{r,r'}^{(c)}(\tau_1-\tau_2)[W^{-1}]_{r',c}(\tau_2).
	\label{equ:hyb_funct}
	\end{equation}
	Here, we have introduced the propagator
	\begin{equation}
		U_{r,r'}^{(c)}(\tau)=\langle\mathcal{T}\varphi_r(\tau)\varphi_{r'}(0)\rangle_{S^{(c)}}^{\rm con}
	\end{equation}
	for the auxiliary fields on the cavity lattice. With this, we have mapped the lattice model to an approximate impurity problem
	\begin{equation}
	S_{HS}^{\rm imp}=S_{0}^c+S_{\varphi\varphi}^{\rm imp}+S_{\varphi\sigma}^{\rm imp}
	\label{equ:imp_HS}
	\end{equation}
	with the linear coupling term
	\begin{equation}
	S_{\varphi\sigma}^{\rm imp}=\,S^c_{\varphi\sigma}+S^{\rm hyb}_1
	=-\int\limits_0^{\beta}d\tau\varphi_c(\tau)[\sigma^1_c(\tau)-h(\tau)]
	\end{equation}
	and the quadratic term
	\begin{equation}
	S_{\varphi\varphi}^{\rm imp}=\,S^c_{\varphi\varphi}+S^{\rm hyb}_2
	=\frac{1}{2}\int\limits_0^{\beta}d\tau\int\limits_0^{\beta}d\tau'\varphi_{c}(\tau)\mathcal{W}^{-1}(\tau-\tau')\varphi_c(\tau').
	\end{equation}
	In the above expression, we have introduced the Weiss field
	\begin{equation}
	\mathcal{W}^{-1}(\tau)=[W^{-1}]_{c,c}(\tau)-\Delta_{\rm hyb}(\tau),
	\end{equation}
	which is one of the central quantities of DMFT.  
	\subsection{Derivation of the self-consistency conditions}
	\label{sec:deriv_sc}
	In the previous section, we have mapped the lattice model to a local impurity problem. However, the impurity action \eqref{equ:imp_HS} contains the fields $h(\tau)$ and $\mathcal{W}(\tau)$, which cannot be 
        evaluated 
	analytically. We will thus express these fields in terms of the corresponding lattice quantities and a local self-energy.
	
	Let us first define the full propagator
	\begin{equation}
	U_{r,r'}(\tau-\tau')=\langle T \varphi_r(\tau)\varphi_{r'}(\tau')\rangle_{S_{HS}}^{\rm con}	
	\end{equation}
	for the auxiliary fields. Introducing the self-energy $\Pi_k$, it is given by the Dyson equation
	\begin{equation}
	U_k^{-1}=W_k^{-1}-\Pi_k.
	\end{equation}
	We may express the propagator for the lattice with the cavity at site $c$ in terms of the full propagator by removing the connection between site $c$ and all other lattice points $r$ and $r'$, i.e.
	\begin{equation}
	U_{r,r'}^{(c)}(i\nu_n)=U_{r,r'}(i\nu_n)-\frac{U_{rc}(i\nu_n)U_{cr'}(i\nu_n)}{U_{cc}(i\nu_n)}.
	\end{equation}
	Here we have divided the second term by $U_{cc}(i\nu_n)$ to avoid double counting. 
	
	Inserting the expression above into the formula for the hybridization function \eqref{equ:hyb_funct} and transforming to $\bm k$-space yields
	\begin{equation}
	\Delta_{\rm hyb}(i\nu_n)=\frac{1}{N}\sum_k E_k^2 U_k - \frac{\left(\frac{1}{N}\sum_k E_k U_k\right)^2}{\frac{1}{N}\sum_k U_k}
	\end{equation}
	with
	\begin{equation}
	E_k=W_k^{-1}-[W^{-1}]_{c,c}.
	\end{equation}
	As it is usually done in DMFT, we assume that the self-energy is purely local, i.e. $\Pi_k=\Pi_{\rm loc}$, and define
	\begin{equation}
	F^{-1}=[W^{-1}]_{c,c}-\Pi_{\rm loc}.
	\end{equation}
	With this we can rewrite the Dyson equation for the full propagator as
	\begin{equation}
	U_k^{-1}=F^{-1}+E_k,
	\end{equation}
	and thus
	\begin{equation}
	\begin{split}
	\frac{1}{N}\sum_k E_k U_k=&\,\frac{1}{N}\sum_k \frac{E_k}{F^{-1}+E_k}=\frac{1}{N}\sum_k\left(\frac{E_k+F^{-1}}{F^{-1}+E_k}-\frac{F^{-1}}{F^{-1}+E_k}\right)\\
	=&\,1-F^{-1}\frac{1}{N}\sum_k U_k
	\end{split}
	\end{equation}
	and
	\begin{equation}
	\begin{split}
	\frac{1}{N}\sum_k E_k^2 U_k=&\,\frac{1}{N}\sum_k E_k\frac{E_k}{F^{-1}+E_k}=\frac{1}{N}\sum_k E_k\left(1-\frac{F^{-1}}{F^{-1}+E_k}\right)\\
	=&\underbrace{\frac{1}{N}\sum_k E_k}_{=0} - F^{-1}\frac{1}{N}\sum_k E_kU_k=-F^{-1}+\left(F^{-1}\right)^2\frac{1}{N}\sum_kU_k
	\end{split}.
	\end{equation}
	Putting everything together we obtain
	\begin{equation}
	\begin{split}
	\mathcal{W}^{-1}(i\nu_n)=&\,[W^{-1}]_{c,c}(i\nu_n)-\Delta(i\nu_n)\\
	=&\,[W^{-1}]_{c,c}(i\nu_n)-F^{-1}(i\nu_n)+[U_{c,c}]^{-1}(i\nu_n)\\
	=&\,[U_{c,c}]^{-1}(i\nu_n)+\Pi_{loc}(i\nu_n).
	\label{equ:Weiss_U}
	\end{split}
	\end{equation}
	With this, we have found an expression for the Weiss field that only contains quantities corresponding to the full lattice without cavity at site $c$.
	
	In the next step, we 
	relate $h(\tau)$ to the Weiss field. For that purpose, we consider the derivative $\frac{\delta\ln\mathcal{Z}}{\delta\varphi_c(\tau)}$. Calculating the partition function from the full lattice action $S_{HS}$ yields
	\begin{equation}
	\frac{\delta\ln\mathcal{Z}}{\delta\varphi_c(\tau)}=\langle \sigma_c^1(\tau)\rangle-\sum_r\int\limits_0^\beta d\tau' [W^{-1}]_{cr}(\tau-\tau')\langle\varphi_r(\tau')\rangle.
	\end{equation}
	On the other hand, evaluating the expression using $S_{HS}^{\rm imp}$ we obtain
	\begin{equation}
	\frac{\delta\ln\mathcal{Z}}{\delta\varphi_c(\tau)}=\langle \sigma_c^1(\tau)\rangle-h(\tau)-\int\limits_0^\beta d\tau'\,\mathcal{W}^{-1}(\tau-\tau')\langle\varphi_c(\tau')\rangle.
	\end{equation}
	Let us assume that the expectation value $\langle\varphi_r(\tau')\rangle$ is time independent and uniform, such that 
	$\langle\varphi_r(\tau')\rangle=\phi$ for all $\tau'$ and $r$.
	Then, defining
	\begin{equation}
	\mathcal{W}^{-1}_0=\mathcal{W}^{-1}(i\nu_n=0)
	\end{equation}
	and
	\begin{equation}
	W_{\rm mf}^{-1}=W_{k=0}^{-1}(i\nu_n=0)
	\end{equation}
	we finally get
	\begin{equation}
	h=\left[W_{mf}^{-1}-\mathcal{W}^{-1}_0\right]\phi.
	\end{equation}

	\subsection{Eliminating the auxiliary field from the impurity problem}
	\label{sec:elim_aux_fields}
	To obtain an impurity action $S^{\rm imp}$ that only depends on the matter degrees of freedom, we integrate out the auxiliary field $\varphi_c$ from $S_{HS}^{\rm imp}$. The path integral
	\begin{equation}
	\int\mathcal{D}[\varphi_0]e^{-S_{HS}^{\rm imp}}=e^{-S^{\rm imp}}
	\end{equation}
	can be solved analytically and yields
	\begin{equation}
	S^{\rm imp}=S^{c}_{0}-\frac{1}{2}\int\limits_0^\beta d\tau\int\limits_0^\beta d\tau'[\sigma_c^1(\tau)-h]\mathcal{W}(\tau-\tau')[\sigma^1_c(\tau')-h].
	\end{equation}
	This can be cast into a slightly different form
	\begin{equation}
	S^{\rm imp}=S^{c}_{0}+S_{{\rm int},1}^{\rm imp}+S_{{\rm int},2}^{\rm imp}+\text{const.}
	\label{equ:S_imp}
	\end{equation}
	with a linear field term 
	\begin{equation}
	S_{{\rm int},1}^{\rm imp}=b\int\limits_0^\beta d\tau \sigma^1_c(\tau),
	\end{equation}
	and the quadratic term 
	\begin{equation}
	S_{{\rm int},2}^{\rm imp}=-\frac{1}{2}\int\limits_0^\beta d\tau\int\limits_0^\beta d\tau'\,\sigma^1_c(\tau)\mathcal{W}(\tau-\tau')\sigma^1_c(\tau'),
	\end{equation}
	where
	\begin{equation}
		b=\mathcal{W}_0\left[W_{mf}^{-1}-\mathcal{W}_0^{-1}\right]\phi.
	\end{equation}
	To eliminate $\phi$, we take into account that the action satisfies the condition
	\begin{equation}
	\left\langle \frac{\delta S_{HS}}{\delta \varphi_r(\tau)} \right\rangle_{S_{HS}}\Bigg\vert_{\varphi_r(\tau)=\phi}=\sum_{r'}\int\limits_0^\beta d\tau' [W^{-1}]_{r,r'}(\tau-\tau')\phi-\langle \sigma^1_r(\tau) \rangle_{S_{HS}}\Bigg\vert_{\varphi_r(\tau)=\phi}=0,
	\end{equation}
	since $\phi$ is a stationary path; therefore,  
	\begin{equation}
	\langle \sigma^1_r(\tau) \rangle_{S_{HS}}\Bigg\vert_{\varphi_r(\tau)=\phi}=\langle\sigma^1\rangle.
	\end{equation}
	With this, we obtain
	\begin{equation}
	\phi=W_{mf}\langle\sigma^1\rangle,
	\end{equation}
	and, thus
	\begin{equation}
	b=\left[\mathcal{W}_0-W_{mf}\right]\langle\sigma^1\rangle.
	\end{equation}

	It is not possible to calculate the local propagator $U_{c,c}$ directly from the impurity action above. However, we may relate the local correlation function 
	\begin{equation}
	\chi_{c,c}=\langle T\sigma^1_c(\tau)\sigma^1_c(\tau')\rangle^\text{con}
	\end{equation}
	to $U_{c,c}$. For that purpose, we define the
	generating functional
	\begin{equation}
	\mathcal{G}[J]=\ln\left\langle \exp\left[\int_0^\beta d\tau\,J(\tau)\varphi_c(\tau)\right]\right\rangle.
	\end{equation}
	Using the impurity action $S_{HS}^{\rm imp}$ with the auxiliary fields, we obtain
	\begin{equation}
	\frac{\delta^2\mathcal{G}[J]}{\delta J(\tau)\delta J(\tau')}\bigg\vert_{J=0}=U_{c,c}(\tau-\tau').
	\end{equation}
	On the other hand, a calculation with $S^{\rm imp}$ yields
	\begin{equation}
	\frac{\delta^2\mathcal{G}[J]}{\delta J(\tau)\delta J(\tau')}\bigg\vert_{J=0}=\mathcal{W}(\tau-\tau')+\int\limits_0^\beta d\tau_1\int\limits_0^\beta d\tau_2 \mathcal{W}(\tau-\tau_1)\chi_{c,c}(\tau_1-\tau_2)\mathcal{W}(\tau_2-\tau')
	\end{equation}
	and thus
	\begin{equation}
	U_{c,c}=\mathcal{W}+\mathcal{W}\chi_{c,c}\mathcal{W}.
	\end{equation}
	Substituting this into Eq.~\eqref{equ:Weiss_U} we obtain
	\begin{equation}
	\Pi_{\rm loc}=\left(1+\chi_{c,c}\mathcal{W}\right)^{-1}\chi_{c,c}.
	\end{equation}
	This allows us to calculate all relevant quantities on the impurity directly from the impurity action \eqref{equ:S_imp}.

	\section{Tail correction}
	\label{sec:tail_app}
	\subsection{High-frequency behavior of the Weiss field}
	To estimate the asymptotic behavior of the Weiss field for $i\nu_n\rightarrow\infty$, we consider an expansion of all relevant quantities in the inverse Matsubara frequency and only keep
	terms up to the order $i\nu_n^{-2}$. As will be discussed in greater detail in App.~\ref{sec:chi_tail}, the local dipole-dipole correlation function decays as
	\begin{equation}
		\chi_{c,c}\sim\frac{c_2}{(i\nu_n)^2}.
		\label{equ:chi_tail}
	\end{equation}
	The interaction matrix can be written as
	\begin{equation}
		W_{\rm k}\sim w_0^{\rm k}+\frac{w_2^{\rm k}}{(i\nu_n)^2}
		\label{equ:W_k_tail}
	\end{equation}
	for $i\nu_n\rightarrow\infty$. The corresponding parameters will be derived in App.~\ref{sec:W_tail}. Moreover, at large Matsubara frequencies we can make the ansatz
	\begin{equation}
		\mathcal{W}\sim w_0+\frac{w_2}{(i\nu_n)^2}
		\label{equ:Weiss_tail}
	\end{equation}
	for the Weiss field.
	Using Eq.~\eqref{equ:self-energy}, we find that the self energy decays as
	\begin{equation}
		\Pi_{\rm loc}=\,\chi_{c,c}-\chi_{c,c}\mathcal{W}\chi_{c,c}+\chi_{c,c}\mathcal{W}\chi_{c,c}\mathcal{W}\chi_{c,c}- \cdots
		\sim\,\frac{c_2}{(i\nu_n)^2}+\mathcal{O}((i\nu_n)^{-4}).
	\end{equation}
	Inserting this into the Dyson equation yields
	\begin{equation}
		U_{\rm k}=\,W_{\rm k}+W_{\rm k}\Pi_{\rm loc}W_{\rm k}+W_{\rm k}\Pi_{\rm loc}W_{\rm k}\Pi_{\rm loc}W_{\rm k}+\cdots
		\sim\,w_0^{\rm k}+\frac{w_2^{\rm k}+(w_0^{\rm k})^2c_2}{(i\nu_n)^2}+\mathcal{O}((i\nu_n)^{-4}),
	\end{equation}
	and, therefore,
	\begin{equation}
		U_{c,c}=\frac{1}{N}\sum_{\rm k}w_0^{\rm k}+\frac{\frac{1}{N}\sum_{\rm k}w_2^{\rm k}+\frac{1}{N}\sum_{\rm k}(w_0^{\rm k})^2c_2}{(i\nu_n)^2}+\mathcal{O}((i\nu_n)^{-4})
	\end{equation}
	for $i\nu_n\rightarrow\infty$. Then Eq.~\eqref{equ:self-energy1} finally yields
	\begin{equation}
		\begin{split}
		\mathcal{W}=&\,U_{c,c}-U_{c,c}\Pi_{\rm loc}U_{c,c}+U_{c,c}\Pi_{\rm loc}U_{c,c}\Pi_{\rm loc}U_{c,c}-...\\
		\sim&\,m_0^{(1)}+\frac{m_2^{(1)}+\left[m_0^{(2)}-\left(m_0^{(1)}\right)^2\right]c_2}{(i\nu_n)^2}+\mathcal{O}((i\nu_n)^{-4}),
		\end{split}
	\end{equation}
	where we have introduced the notation
	\begin{align}
		m_0^{(1)}&=\frac{1}{N}\sum_{\rm k}w_0^{\rm k},\\
		m_0^{(2)}&=\frac{1}{N}\sum_{\rm k}\left(w_0^{\rm k}\right)^2,\\
		m_2^{(1)}&=\frac{1}{N}\sum_{\rm k}w_2^{\rm k}.
	\end{align}
	With this, we can identify the parameters in Eq.~\eqref{equ:Weiss_tail} for the high-frequency tail of the Weiss field as
	\begin{equation}
		w_0=m_0^{(1)}
	\end{equation}
	and
	\begin{equation}
		w_2=m_2^{(1)}+\left[m_0^{(2)}-\left(m_0^{(1)}\right)^2\right]c_2.
	\end{equation}
	\subsection{High-frequency behavior  of the local dipole-dipole correlation function}
	\label{sec:chi_tail}
	In the following, we derive the approximate expression for the local dipole-dipole correlation function at high Matsubara frequencies given in Eq.~\eqref{equ:chi_tail}, and determine the coefficient $c_2$. For that purpose, we make use of the spectral representation. The spectral function is defined as
	\begin{equation}
		A(\omega)=-\frac{1}{\pi}{\rm Im}\left\{\chi_{c,c}(\omega)\right\}.
	\end{equation}
	With this, the local dipole-dipole correlation function in the Matsubara representation is formally given by
	\begin{equation}
		\begin{split}
		\chi_{c,c}(i\nu_n)
		=&\int\limits_0^{\beta}d\omega A(\omega)\left[\frac{1}{i\nu_n-\omega}-\frac{1}{i\nu_n+\omega}\right]\\
		=&\int\limits_0^{\beta}d\omega A(\omega)\frac{2\omega}{(i\nu_n)^2}\frac{1}{1-(\omega/i\nu_n)^2}\\
		=&\frac{1}{(i\nu_n)^2}\int\limits_0^{\beta}2\omega A(\omega)+\mathcal{O}((i\nu_n)^{-4}).
		\end{split}
	\end{equation}
	To obtain the last line, we have made a Taylor expansion in $1/(i\nu_n)$. At large Matsubara frequencies, we can neglect all terms beyond the second order and, thus, we find
	\begin{equation}
		c_2=\int\limits_0^{\beta}d\omega 2\omega A(\omega).
		\label{equ:c2_1}
	\end{equation}

	On the other hand, the imaginary time representation of the local dipole-dipole correlation function can be rewritten in terms of the spectral function as
	\begin{equation}
		\chi_{c,c}(\tau)=\int\limits_0^{\beta}d\omega A(\omega) \frac{\cosh[\omega(\tau-\beta/2)]}{\sinh[\omega\beta/2]}.
	\end{equation}
	Taking the derivative with respect to time yields
	\begin{equation}
		\partial_{\tau}\chi_{c,c}(\tau)=\int\limits_0^{\beta}d\omega A(\omega)\omega\frac{\sinh[\omega(\tau-\beta/2)]}{\sinh[\omega\beta/2]}
	\end{equation}
	and, thus,
	\begin{equation}
		\partial_{\tau}\chi_{c,c}(\tau)\bigg\vert_{\tau=0}=-\int\limits_0^{\beta}d\omega\omega A(\omega).
	\end{equation}
	Comparing this to Eq.~\eqref{equ:c2_1}, it can be seen immediately that
	\begin{equation}
		c_2=-2\partial_{\tau}\chi_{c,c}(\tau)\bigg\vert_{\tau=0}.
	\end{equation}

	In order to calculate the derivative, we recall that
	\begin{equation}
		\partial_{\tau}\chi_{c,c}(\tau)=\partial_{\tau}\langle\mathcal{T}\hat{\sigma}_c^x(\tau)\hat{\sigma}_c^x(0)\rangle,
	\end{equation}
	and, therefore, for $\tau>0$
	\begin{equation}
		\partial_{\tau}\chi_{c,c}(\tau)=\langle[\partial_{\tau}\hat{\sigma}_c^x(\tau)]\hat{\sigma}_c^x(0)\rangle.
	\end{equation}
	In the modified Heisenberg picture, the equation of motion for the $x$-component of the Pauli operator reads
	\begin{equation}
		\partial_{\tau}\hat{\sigma}_c^x(\tau)
		=\left[\hat{H},\hat{\sigma}_c^x(\tau)\right]
		=\sum_r\frac{\Delta}{2}[\hat{\sigma}_r^3(\tau),\hat{\sigma}_c^x(\tau)]
		=\Delta i\hat{\sigma}_c^y(\tau).
	\end{equation}
	With this,
	\begin{equation}
		\partial_{\tau}\chi_{c,c}(\tau)\bigg\vert_{\tau=0}=\Delta i\langle\hat{\sigma}_c^y\hat{\sigma}_c^x\rangle=\Delta\langle\hat{\sigma}_c^z\rangle,
	\end{equation}
	which finally yields
	\begin{equation}
		c_2=-2\Delta\langle\hat{\sigma}_c^z\rangle.
	\end{equation}

	\subsection{High-frequency behavior of the dipole-dipole interaction vertex}
	\label{sec:W_tail}
	In Eq.~\eqref{equ:W_k_tail} we have introduced an approximation for the interaction matrix at large Matsubara frequencies. The coefficients $w_0^{\rm k}$ and $w_2^{\rm k}$ can be obtained from an expansion of $W_{\rm k}(i\nu_n)$ in $1/(i\nu_n)$. This yields
	\begin{equation}
		\begin{split}
		W_{\rm k}(i\nu_n)
		=&\frac{\alpha}{2}\left[\cos(k_y)+\cos(k_z)\right]
		-|g_{\bm k}|^2\frac{\nu_n^2}{\nu_n^2+\omega_{\bm k}^2}\\
		=&\frac{\alpha}{2}\left[\cos(k_y)+\cos(k_z)\right]-|g_{\bm k}|^2\frac{1}{1+(\omega_{\bm k}/\nu_n)^2}\\
		=&\frac{\alpha}{2}\left[\cos(k_y)+\cos(k_z)\right]-|g_{\bm k}|^2-\frac{|g_{\bm k}|^2\omega_{\bm k}^2}{(i\nu_n)^2}+\mathcal{O}((i\nu_n)^{-4})
		\end{split}
	\end{equation}
	and, thus,
	\begin{align}
		w_0^{\bm k}=&\frac{\alpha}{2}\left[\cos(k_y)+\cos(k_z)\right]-|g_{\bm k}|^2, \\
		w_2^{\bm k}=&-|g_{\bm k}|^2\omega_{\bm k}^2.
	\end{align}
\end{document}